\documentclass[manuscript]{acmart}  

\AtBeginDocument{%
  \providecommand\BibTeX{{%
    \normalfont B\kern-0.5em{\scshape i\kern-0.25em b}\kern-0.8em\TeX}}}

\setcopyright{acmcopyright}
\copyrightyear{2018}
\acmYear{2018}
\acmDOI{XXXXXXX.XXXXXXX}

\acmConference[Conference acronym 'XX]{Make sure to enter the correct
  conference title from your rights confirmation emai}{June 03--05,
  2018}{Woodstock, NY}
\acmPrice{15.00}
\acmISBN{978-1-4503-XXXX-X/18/06}

\usepackage{tabularx}
\usepackage{subfigure}
\usepackage{multirow}
\usepackage{booktabs}
\usepackage{indentfirst} 
\usepackage{fancyhdr}
\usepackage{enumitem}
\usepackage{algorithm}
\usepackage{algorithmic}
\usepackage{makecell}
\usepackage{xcolor}
\usepackage{bbm}

\begin{document}

\title{SoREX: Towards Self-Explainable Social Recommendation with Relevant Ego-Path Extraction}

\author{Hanze Guo}
\authornote{Both authors contributed equally to this research.}
\affiliation{
  \institution{Gaoling School of Artificial Intelligence, Renmin University of China}
  \city{Beijing}
  \country{China}
  }
\email{ghz@ruc.edu.cn}

\author{Yijun Ma}
\authornotemark[1]
\affiliation{
  \institution{Gaoling School of Artificial Intelligence, Renmin University of China}
  \city{Beijing}
  \country{China}
  }
\email{mayj_hedgehog@ruc.edu.cn}

\author{Xiao Zhou}
\authornote{Corresponding author}
\affiliation{%
  \institution{Gaoling School of Artificial Intelligence, Renmin University of China}
  \city{Beijing}\country{China}}
\additionalaffiliation{%
  \institution{Beijing Key Laboratory of Research on Large Models and Intelligent Governance}
  \city{Beijing}\country{China}}
\additionalaffiliation{%
  \institution{Engineering Research Center of Next-Generation Intelligent Search and Recommendation, MOE}
  \city{Beijing}\country{China}}
\email{xiaozhou@ruc.edu.cn}

\renewcommand{\shortauthors}{Guo, et al.}

\begin{abstract}

Social recommendation has been proven effective in addressing data sparsity in user-item interaction modeling by leveraging social networks. The recent integration of Graph Neural Networks (GNNs) has further enhanced prediction accuracy in contemporary social recommendation algorithms. However, many GNN-based approaches in social recommendation lack the ability to furnish meaningful explanations for their predictions. In this study, we confront this challenge by introducing SoREX, a self-explanatory GNN-based social recommendation framework. 
SoREX adopts a two-tower framework enhanced by friend recommendation, independently modeling social relations and user–item interactions, while jointly optimizing an auxiliary task to reinforce social signals.
To offer explanations, we propose a novel ego-path extraction approach. This method involves transforming the ego-net of a target user into a collection of multi-hop ego-paths, from which we extract factor-specific and candidate-aware ego-path subsets as explanations. This process facilitates the summarization of detailed comparative explanations among different candidate items through intricate substructure analysis. Furthermore, we conduct explanation re-aggregation to explicitly correlate explanations with downstream predictions, imbuing our framework with inherent self-explainability. Comprehensive experiments conducted on four widely adopted benchmark datasets validate the effectiveness of SoREX in predictive accuracy. Additionally, qualitative and quantitative analyses confirm the efficacy of the extracted explanations in SoREX. Our code and data are
available at https://github.com/antman9914/SoREX.
\end{abstract}




%


\maketitle

\section{Introduction}

Recommender systems have become a common choice for alleviating information overload after the prosperity of the Internet~\cite{cfevaluating,medretrieval}. However, they always suffer from the sparse interactions between users and items. According to social correlation theories~\cite{cialdini2004social}, it is believed that users' preferences can be influenced by their social relationships. With the proliferation of online social platforms, social recommendation has been designed to improve user-item interaction modeling, addressing data sparsity and cold start problem with the aid of user-user social regularizations~\cite{socialmf,trustmf}.

Recently, Graph Neural Networks (GNNs)~\cite{lightgcn,tricolore} have been widely adopted in recommender systems owing to their robust ability to model structured data. Notably, contemporary social recommendation algorithms~\cite{diffnet,mhcn,design,cgcl,tail} have embraced GNNs to harness high-order social context and collaborative information simultaneously. Despite the significant accuracy enhancements brought about by GNN-based methods, they often fall short in providing meaningful explanations for their predictions. 

However, the explainability of predictions made by social recommender systems is of paramount importance for both users and service providers. For users, explainability serves as a foundation for fostering engagement and trust in the system~\cite{zhang2014explicit}, enabling them to make more informed decisions~\cite{sarwar2001item}. Online A/B testing on e-commerce platforms has demonstrated that providing explanations can improve click-through rates in real-world commercial settings~\cite{zhang2014explicit}. For service providers, explainability is equally critical. High-profile incidents, such as YouTube’s loss of advertising revenue due to the opaque nature of its recommendation algorithms~\cite{hern2017youtube}, and Amazon’s discontinuation of a biased recruitment system~\cite{dastin2018amazon}, underscore the potential risks and consequences of lacking transparency. Therefore, enhancing the explainability of recommendation models not only improves system transparency but also aids in uncovering systematic patterns, thereby deepening our understanding of underlying network characteristics~\cite{zhang2020explainable}.

Current research on explainable recommendation systems often draws on user reviews~\cite{narre,mter,fact} and knowledge graphs (KGs)~\cite{kprn,KG-metapath,KGR-rl}. Review-based methods focus on generating coherent and persuasive explanations for users, but they heavily depend on the availability and quality of textual reviews~\cite{narre}. In many scenarios, such reviews may be sparse, biased, or even absent, especially for newly added items or users, which greatly limits the generalizability of these methods. On the other hand, KG-based methods aim to improve transparency by leveraging semantic paths within well-constructed knowledge graphs~\cite{KG-metapath}. However, such graphs are often unavailable or incomplete in practice, particularly in social networks or user–item interaction graphs, which restricts the applicability of KG-based approaches.

To overcome the limitations of existing explanation methods, we emphasize substructure mining~\cite{gnnexplainer,conpi}—a widely used graph-based interpretability technique that extracts the most contextually relevant and interpretable subgraphs. This approach helps uncover complex, factor-specific patterns essential for understanding the reasoning behind recommendation rankings. In this work, we leverage the structural properties of user–item interaction graphs and social networks to enhance the transparency of social recommender systems, enabling more general and broadly applicable explanations. However, most substructure mining methods are post hoc~\cite{gnnexplainer,pgexplainer,rgexplainer}, generating explanations only after model training. Such approaches are prone to distribution shifts~\cite{mixup-explainer,post-hoc-consistent}, which can compromise explanation fidelity. As a result, recent efforts have focused on developing self-explainable GNNs~\cite{conpi,gsat,disc,pxgnn} that generate explanatory subgraphs as part of the prediction process. Nevertheless, existing methods primarily target graph-level classification and often fail to capture the complex, node-pair-specific substructures required for recommendation tasks. Moreover, explanations in recommendation typically answer the question, “Why should this user choose this item?”—yet in ranking-based settings, more persuasive explanations may arise from comparative reasoning, i.e., “Why should this item be ranked higher than others?” While~\cite{compare-explain} attempts to address this question, it relies heavily on textual reviews and does not explicitly improve model transparency.

\begin{figure}
    \centering
    \includegraphics[width=0.6\linewidth]{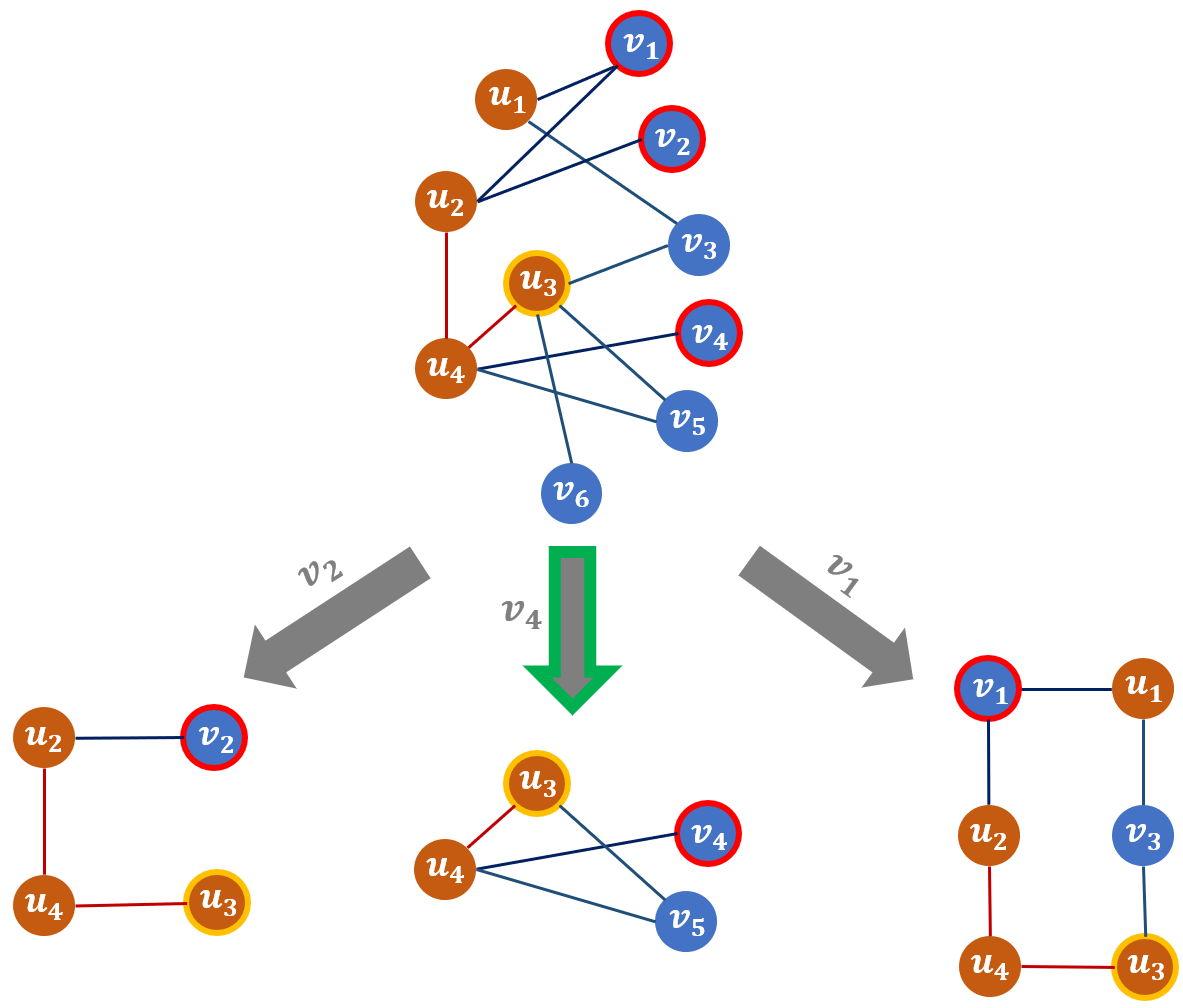}
    \caption{A toy example illustrating a substructure-based comparative explanation for recommending item $v_4$ to user $u_3$. We enumerate all meta-paths of length no longer than 3 from $u_3$ to items $v_{1}$, $v_{2}$, and $v_{4}$. While $v_{1}$ and $v_{4}$ each have two paths and $v_{2}$ only one, $v_{4}$ is more relevant due to a shorter, socially meaningful path via direct friend $u_{4}$. This highlights how structural signals alone can yield accurate and interpretable recommendations.}
    \label{fig:intro_toy}
\end{figure}

Therefore, two major challenges remain: \textbf{(1) Pair-wise complexity}: While graph-based explanation methods excel at providing node-level interpretability, they often struggle to capture the pair-wise interpretability essential for recommendation tasks, where understanding the relationship between specific user-item pairs is crucial; \textbf{(2) Comparative explanations}: Existing methods rarely provide effective comparative insights between different items in the recommendation ranking, limiting their ability to answer not just ‘why this item?’ but also ‘why this item over others?’, which is vital for persuasive and transparent recommendations. Nevertheless, our proposed method can effectively overcome both of these challenges by extracting and aggregating candidate-aware ego-path subgraphs for each user–item pair, and supporting direct comparison among candidates for transparent, comparative ranking explanations. As illustrated in Fig.~\ref{fig:intro_toy}, we can \emph{solely} leverage
the structural user–item and social graph—without reviews, textual KG facts
or other hard-to-obtain signals—by jointly evaluating the number and the
quality of meta-paths of length~$\le 3$; in this toy case, this structural
criterion already lifts $v_{4}$ to the top of the ranking.
Such \emph{comparative}, structure-based explanations can be surfaced to
both users and developers, improving recommendation accuracy while
clarifying the model’s decision process. 

To fill the comparative explainability gap in GNN-based social recommendation, we propose a novel framework named \textit{SoREX} (\textit{So}cial \textit{R}ecommendation based on relevant \textit{E}go-path e\textit{X}traction). SoREX is designed to be self-explainable, enabling the exploration of candidate-aware complex substructures and comparative relationships among candidate items from different perspectives. 
It is built on a two-tower architecture: a social influence-aware social tower and an interaction tower, each equipped with separate GNN encoders to learn from the social network and user-item interaction graph respectively. The towers independently conduct user/item embedding and candidate item ranking, with final rankings obtained by fusing both tower predictions. 
Such design can independently model social and interaction factors, laying foundation for factor-specific explanations.
Multi-task learning is further employed during training, where a friend-recommendation auxiliary task is attached to the social tower to capture more reliable friend relations.

For clarity, we denote users' multi-hop neighborhoods as their \textit{ego-nets}. Since GNNs can only perceive the ego-net of the target user, we propose to extract a subgraph of the ego-net relevant to each individual candidate item and factor, serving as a candidate-aware and factor-specific explanation. To enable complex substructure investigation and provide detailed comparison, we aim to extract dense explanatory subgraphs instead of sparse ones, as suggested by~\cite{lri,gsat}. We achieve this by transforming the ego-net into a set of \textit{ego-paths}, shared by both towers, representing all multi-hop paths on the joint graph of the social network and user-item interaction graph originating from the target user. 
Random walk sampling ensures memory efficiency during ego-path generation.
Next, we compute similarities between the given candidate and all observed ego-paths in each tower. Transforming these similarities into ego-path sampling probabilities, we obtain a subset of ego-paths relevant to each candidate and factor. Finally, explanation re-aggregation aggregates information from the sampled ego-path subsets into the final user representations used for downstream prediction, emphasizing candidate-aware and factor-specific neighborhood information. Ego-paths, forming motifs through interweaving, enable complex substructure investigation. Additionally, each candidate is assigned factor-specific explanation graphs with different structures and similarity distributions, naturally forming comparative relationships and explanations among candidate items.

In summary, our contributions are highlighted as follows:
\begin{itemize}
     \item We introduce a two-tower GNN-based social recommendation framework that independently models user preferences from social and user-item interaction perspectives.
     \item We devise a novel explanation extraction strategy that samples factor-specific and candidate-aware subsets of multi-hop ego-paths for each candidate item. This enables detailed comparative explanations for ranking predictions and facilitates high-level substructure analysis.
     \item We propose explanation re-aggregation to connect explanatory ego-path subsets to predictions, making our framework self-explainable. To the best of our knowledge, we are the first to address the comparative explainability gap in GNN-based social recommendation.
     \item Empirical experiments on four benchmark datasets substantiate the superiority of SoREX in accuracy. Additionally, we qualitatively and quantitatively demonstrate the explainability of our method.
\end{itemize}

\section{Related Work}

\subsection{Social Recommendation}

Recommender systems have been widely explored as an effective approach for modeling user–item interactions~\cite{lightgcn,ada,topic}. 
However, the sparsity of user-item interaction often limits their performance. To alleviate this issue, social recommendation is proposed to incorporate social networks and enrich user–item interaction modeling. Early studies~\cite{socialmf,sr,cnsr,trustsvd,trustmf} usually use social network for embedding regularization. 

Recent works introduce prospering GNNs into social recommendation due to their strong capability in modeling relational data. GraphRec~\cite{graphrec} is the pioneer to design social recommender with attention-based GNN. To model the recursive influence diffusion process, Diffnet~\cite{diffnet} and  Diffnet++~\cite{diffnet++} adopt a multi-layer influence propagation architecture to learn the evolution of user preference. MHCN~\cite{mhcn} proposes a multi-channel hypergraph convolutional network to explicitly leverage high-order user relations. DESIGN~\cite{design} leverages multiple GNN encoders to learn from different factors, fusing their knowledge via knowledge distillation. SocialLGN~\cite{sociallgn} conducts message propagation in both social network and interaction graph in each GNN layer for knowledge fusion. 
ESRF~\cite{esrf} devises a GCN-based deep adversarial social recommendation framework. Jiang et al.~\cite{share} identify the low preference homophily among socially connected users, and propose SHaRe to address the issue via social graph rewiring. 
Recent works also integrate self-supervised learning~\cite{sept,derec,cgcl} and graph denoising~\cite{gbsr,recdiff,gdssl} into GNN-based social recommendation. 

Although these methods significantly improve accuracy, they largely overlook interpretability, leaving users unclear about the reasons behind recommendations. This lack of explanation limits transparency, which is critical for real-world applications. To address this, SoREX is designed as a self-explainable GNN-based social recommender that highlights user preferences and social interactions, improving transparency through substructure-aware and comparative explanations.

\subsection{Explainable Recommendation}

Most explainable recommendation solutions generate explanations based on item profiles, user reviews and knowledge graphs (KG). Profile-based and review-based methods aim to help users understand their decisions via retrieved or generated textual information. Du et al.~\cite{ecf} proposed a taste cluster based self-explainable collaborative filter method, using item tags for cluster description. As for review-based methods, the retrieval-based methods aim to select review text that matches prediction via attention~\cite{narre}, reinforcement learning~\cite{rl-extract} and other feasible means. A recent work~\cite{compare-explain} also raises the conception of comparative explanation, but it is also a review-based method. On the other hand, generation-based methods aim to identify important aspects of interacted items~\cite{mter,fact} or directly apply text generation techniques~\cite{nrt,saer} to synthesize explanations. KG based methods aim to improve the transparency of recommenders. They mostly employ multi-hop path reasoning on KG , and then apply the optimal paths as explanations.
The difference among them mainly lies in the reasoning strategy, such as meta-path template based~\cite{KG-metapath}, LSTM based~\cite{kprn} and reinforcement learning based~\cite{KGR-rl,KG-rl-2} methods.

Existing profile-based, review-based, and knowledge graph (KG)-based explanation methods rely heavily on external data, such as user reviews or structured semantic triples. However, such data is often difficult to obtain, incomplete, or unavailable, especially in social recommendation scenarios involving new users or items. This limits their applicability in practice. While some prior work~\cite{sr-explain} introduces socially-aware explanations, it still fundamentally depends on reviews. In contrast, \textsc{SoREX} enhances transparency by leveraging only the intrinsic structure of user--item and user--user interactions, without requiring any auxiliary information. Furthermore, most KG-based methods focus on semantic reasoning and are ill-suited for modeling social influence. Our approach fills this gap by introducing a graph-based, self-explainable framework for social recommendation that supports comparative explanations based on mined substructures, thereby improving both interpretability and applicability in real-world settings.

\subsection{Explainable GNN}

The lack of interpretability of GNNs promote the development of GNN explainers. Early GNN explainers are post-hoc, generating posterior substructure-based explanations for trained GNNs. For example, GNNExplainer~\cite{gnnexplainer} and PGExplainer~\cite{pgexplainer} generate edge masks or node feature masks to find the significant subgraphs;
RGExplainer~\cite{rgexplainer} searches for relevant subgraphs via reinforcement learning; 
XGNN~\cite{xgnn} trains a graph generator to generate explanation graphs; 
K-FactExplainer~\cite{kfactexplainer} proposes a factorized explainer to reflect one-to-many relationships between labels and explanatory substructures. 
However, recent works~\cite{post-hoc-consistent, mixup-explainer} prove that post-hoc explainers suffer from distribution shift between explanations and predictions. 

Therefore, efforts have been made to develop self-explainable GNNs. SE-GNN~\cite{se-gnn} is one of the pioneers, using neighbors with the same label as the input node as explanation. ProtGNN~\cite{protgnn} and PxGNN~\cite{pxgnn} identifies representative graph patterns via prototype learning. ConPI~\cite{conpi} and ILP-GNN~\cite{ilp-gnn} model similarity between neighbor sets of given node pair and infer the existence of link between them, which is explained by similar neighbors or neighbor pairs. GIB~\cite{gib}, GSAT~\cite{gsat} and LRI~\cite{lri} all refer to information bottleneck (IB) theory. GIB proposes to train an explanatory graph generator with IB, while GSAT and LRI inject learnable randomness into GNN and sample dense explanatory graphs without spurious correlations. Some recent works have proposed further extensions to IB based methods. For example, PGIB~\cite{protgib} combines prototype learning with IB theory. There is also a group of causal learning based self-explainable GNNs~\cite{cal,disc} proposing to disentangle causal subgraphs from biased graph as explanations; Besides, a recent work GraphChef~\cite{graphchef} proposes to integrate decision tree into the message passing framework of GNNs to achieve self-explainability.

Although some existing GNN-based explainable methods explore complex motifs for explanation, they are primarily designed for classification tasks and lack the ability to extract node pair aware substructures or provide comparative explanations, both of which are essential for ranking-based recommendation. In contrast, \textsc{SoREX} focuses on intrinsic user–item and user–user relationships, making it especially effective for social recommendation. By integrating self-explainable mechanisms that leverage social context and interaction patterns, \textsc{SoREX} fills a key gap in current explainable GNN research. Unlike traditional approaches, it delivers comparative explanations tailored to ranking tasks and does so without relying on external data, enhancing both transparency and practical applicability in real-world systems.

\section{Problem formulation}
\label{sec:preliminary}

\subsection{Graph Based Social Recommendation} 
We denote user set and item set as $\mathcal{U}=\{u_1, u_2, ..., u_m\}$ and $\mathcal{V}=\{v_1, v_2, ..., v_n\}$ respectively, where $m$ is the number of users and $n$ is the number of items. $\mathcal{G}_r=(\mathcal{U} \cup \mathcal{V}, \mathcal{E}_r)$ denotes the user-item bipartite, where $(u_i, v_j)\in \mathcal{E}_r$ indicates that an interaction exists between user $u_i$ and item $v_j$. $\mathcal{G}_s=(\mathcal{U}, \mathcal{E}_s)$ denotes the user-user social graph, where $(u_i, u_j) \in \mathcal{E}_s$ indicates that user $u_i$ and $u_j$ have a social relation. We denote the joint graph of $\mathcal{G}_r$ and $\mathcal{G}_s$ as $\mathcal{G}=(\mathcal{U}\cup\mathcal{V}, \mathcal{E}_r\cup\mathcal{E}_s)$. Let user-item interaction matrix and user-user social relation matrix be $\mathbf{R}\in \mathbb{R}^{m\times n}$ and $\mathbf{S} \in \mathbb{R}^{m\times m}$, where $\mathbf{R}_{ij}=1$ if $(u_i, v_j)\in\mathcal{E}_r$, and $\mathbf{S}_{ij}=1$ if $(u_i, u_j) \in \mathcal{E}_s$. The adjacency matrix $\mathbf{A}^R$ of undirected user-item bipartite and adjacency matrix $\mathbf{A}$ of $\mathcal{G}$ can be defined as Eq. \ref{eq:preliminary}:

\begin{equation}
    \mathbf{A}^R = 
    \begin{bmatrix}
    \mathbf{0} & \mathbf{R} \\
    \mathbf{R}^T & \mathbf{0}
    \end{bmatrix}, 
    \mathbf{A} = 
    \begin{bmatrix}
        \mathbf{S} & \mathbf{R} \\
        \mathbf{R}^T & \mathbf{0}
    \end{bmatrix}.
    \label{eq:preliminary}
\end{equation}

\noindent We formulate the graph based social recommendation problem as follows:

\textit{Definition 1 (Graph based social recommendation):} Given $\mathcal{G}_r$ and $\mathcal{G}_s$, the objective of social recommendation is to predict the missing links in $\mathcal{G}_r$, suggesting the top-$K$ disconnected items that the target users are most likely to interact with.

\subsection{Self-Explainable Social Recommendation}
Generally, self-explainable graph learning methods generate a subgraph for both explanation and downstream prediction. We further treat the explanation graph as an emphasis over relevant subgraph of ego-net in this work to alleviate information loss. 

We first define ego-net as follows:

\textit{Definition 2 (Ego-net):} the $k$-hop ego-net of user $u_i$ is defined as the $k$-hop neighborhood of $u_i$ in $\mathcal{G}$, denoted as $\mathcal{G}_{ego}(i,k)$.

Given ego-net $\mathcal{G}_{ego}(i,k)$, we can represent it with the set of all possible paths originating from $u_i$ no longer than k. By removing redundant information, we derive the definition of ego-path:

\textit{Definition 3 (Ego-path):} Given $u_i$ origined k-hop path $\hat{w}_t(i,k)=\{u_i\rightarrow q_1\rightarrow ... \rightarrow q_{k}\}$, by removing the source node $u_i$ and repetitive nodes, we obtain an ego-path $w_t(i,k)=\{q_1\rightarrow ... \rightarrow q_{k}\}$ no longer than $k-1$.

\begin{figure}
    \centering
    \includegraphics[width=0.6\linewidth]{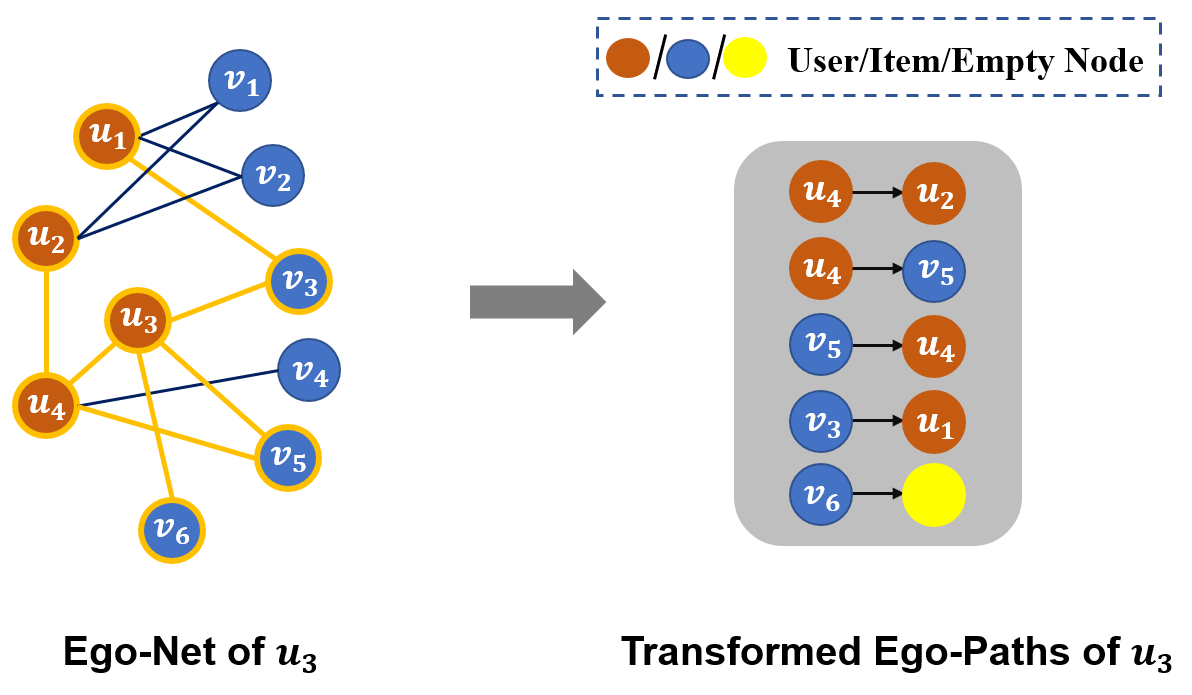}
    \caption{Example of ego-net and ego-path. The 2-hop ego-net of user $u_3$ is highlighted on the left, while the ego-net is transformed into a set of 2-hop ego-paths on the right. Given the length of the ego-path passing by $v_6$ is less than 2, to meet the required length, it is padded with an empty node.}
    \label{fig:intro}
\end{figure}

The complete set of $u_i$-origined ego-path can be denoted as $\mathcal{W}_{ego}(i,k)=\{w_t(i, k)\}_{t=1}^{T}$, where $T$ is the amount of ego-paths.
The ego-paths shorter than $k-1$ are padded to the required length with an empty node with zero vector embedding. Figure \ref{fig:intro} presents a toy example of the relationship between ego-net and ego-path. The yellow empty node indicates padding when there are not enough neighbors, and does not carry any information. We finally define self-explainable social recommendation as follows:

\textit{Definition 4 (Self-explainable social recommendation):}
Given ego-path set $\mathcal{W}_{ego}(i,k)$ of user $u_i$, for each disconnected item $v_j$, we extract candidate-aware ego-path subset $\tilde{\mathcal{W}}_{ego}(i,k,j)$, which is expected to be relevant with $v_j$ and able to explain the ranking of $v_j$. We then predict missing interaction links for $u_i$ based on $\{\tilde{\mathcal{W}}_{ego}(i,k,j)\}_{j=1}^{n_c}, \mathcal{G}_r$ and $\mathcal{G}_s$, where $n_c$ is the amount of disconnected items with $u_i$. For clarity, we simplify notation $\mathcal{G}_{ego}(i,k),\mathcal{W}_{ego}(i,k),\tilde{\mathcal{W}}_{ego}(i,k,j)$, $w_t(i,k)$ as $\mathcal{G}_{ego}, \mathcal{W}_{ego}, \tilde{\mathcal{W}}_{ego}(j), w_t$ in the following text respectively.

\subsection{Friend Recommendation}

Friend recommendation is critical for shaping and facilitating online social networks~\cite{grafrank,ssnet}, helping users to connect with each other. We adopt it as an auxiliary task in SoREX. We can formulate friend recommendation problem as follows:

\textit{Definition 5 (Friend recommendation)}: Given social network $\mathcal{G}_s$, the objective of friend recommendation is to predict missing links in $\mathcal{G}_s$, suggesting top-$K$ disconnected users that the target users are most likely to make friend with.

\begin{table}[h]
  \caption{Summary of key notations used in \textsc{SoREX}.}
  \label{tb:notation}
  \begin{tabularx}{\textwidth}{p{0.3\textwidth}X}
    \Xhline{1pt}
    \multicolumn{2}{l}{\textbf{(A) Data \& Graph Structures}}\\ \Xhline{0.5pt}
    $\mathcal{U}, \mathcal{V}$ & Sets of users and items; $|\mathcal{U}|=m$, $|\mathcal{V}|=n$ \\
    $\mathcal{G}_r=(\mathcal{U}\cup\mathcal{V},\mathcal{E}_r)$ & User–item interaction bipartite graph \\
    $\mathcal{G}_s=(\mathcal{U},\mathcal{E}_s)$ & User–user social graph \\
    $\mathcal{G}=(\mathcal{U}\cup\mathcal{V},\mathcal{E}_r\cup\mathcal{E}_s)$ & Joint graph of $\mathcal{G}_r$ and $\mathcal{G}_s$ \\
    $\mathbf{R}\!\in\!\mathbb{R}^{m\times n}$ & Adjacency / rating matrix of $\mathcal{G}_r$ ($\mathbf{R}_{ij}=1$ if $(u_i,v_j)\!\in\!\mathcal{E}_r$) \\
    $\mathbf{S}\!\in\!\mathbb{R}^{m\times m}$ & Social adjacency matrix ($\mathbf{S}_{ij}=1$ if $(u_i,u_j)\!\in\!\mathcal{E}_s$) \\
    $\mathbf{A}^R,\;\mathbf{A}$ & Block-adjacency matrices of $\mathcal{G}_r$ and $\mathcal{G}$ (Eq.~\ref{eq:preliminary}) \\
    $\mathcal{G}_{ego}(i,k)$ & $k$-hop ego-net of user $u_i$ \\
    $\mathcal{W}_{ego}(i,k)$ & Set of all ego-paths originated from $u_i$ (Def.~3) \\[2pt]
    \Xhline{0.5pt}
    \multicolumn{2}{l}{\textbf{(B) Embeddings \& Representations}}\\ \Xhline{0.5pt}
    $\mathbf{E}^r,\;\mathbf{E}^s$ & ID-embedding matrices for interaction tower / social tower \\
    $\mathbf{e}_q^r,\;\mathbf{e}_q^s$ & ID embedding of node $q$ in the two towers \\
    $\mathbf{h}_i^r,\;\mathbf{c}_j^r$ & GNN-encoded user / item embeddings in interaction tower \\
    $\mathbf{h}_i^s$ & GNN-encoded user embedding in social tower \\
    $\tilde{\mathbf{h}}_j^s$ & Aggregated item representation in social tower (Eq.~\ref{eq:prob-soc}) \\
    $\hat{\mathbf{h}}_i^r,\;\hat{\mathbf{h}}_i^s$ & Explanation-re-aggregated final user embeddings (Eq.~\ref{eq:emphasis_final_emb}) \\[2pt]
    \Xhline{0.5pt}
    \multicolumn{2}{l}{\textbf{(C) Sampling \& Explanation Variables}}\\ \Xhline{0.5pt}
    $w_t$ & One ego-path; $q\!\in\!w_t$ denotes nodes on the path \\
    $\hat{\mathcal{W}}_{ego}$ & Random-walk sampled path pool (size $n_w$) \\
    $\tilde{\mathcal{W}}_{ego}^{r}(j),\;\tilde{\mathcal{W}}_{ego}^{s}(j)$ & Candidate-aware ego-path subsets in two towers \\
    $p_t^{r*}(v_j),\;p_t^{s*}(v_j)$ & Raw path–item similarities (Eq.~\ref{eq:prob-inter},\ref{eq:prob-soc}) \\
    $p_t^{r}(v_j),\;p_t^{s}(v_j)$ & Normalised sampling probabilities (Eq.~\ref{eq:path-sim}) \\
    $\beta_t^{r},\;\beta_t^{s}$ & Bernoulli variables for path sampling  \\
    $\alpha_{qj}^{r},\;\alpha_{qj}^{s}$ & Attention weights in explanation re-aggregation \\[2pt]
    \Xhline{0.5pt}
    \multicolumn{2}{l}{\textbf{(D) Scores, Losses \& Hyper-parameters}}\\ \Xhline{0.5pt}
    $g_r(\!\cdot\!),\;g_s(\!\cdot\!),\;g(\!\cdot\!)$ & Scoring functions of interaction tower, social tower, fused output \\
    $f(u_i,u_q)$ & Friend-recommendation score between users $u_i$ and $u_q$ \\
    $\mathcal{L}_{main},\;\mathcal{L}_s$ & BPR loss for item ranking; BPR loss for friend recommendation \\
    $\gamma,\;\lambda$ & Loss-weight of auxiliary task; $L_2$-regularisation strength \\
    $k_1,k_2$ & GNN propagation layers in interaction / social tower \\
    $k$ & Hop number of ego-net / max walk length \\
    $n_w$ & Number of random walks sampled per user \\
    $d$ & Dimension of embeddings \\
    \Xhline{1pt}
  \end{tabularx}
\end{table}

\section{Methodology}

In this section, we present the methodology behind our proposed SoREX framework. We begin by providing an overview of SoREX and its architecture. Then, we explain the key components in detail, starting with the basic two-tower framework and moving on to the ego-path based explanation process, explanation re-aggregation, and multi-task training. Finally, we conclude with a computational complexity analysis.

\subsection{Overview}

\begin{figure*}
    \centering
    \includegraphics[width=0.85\linewidth]{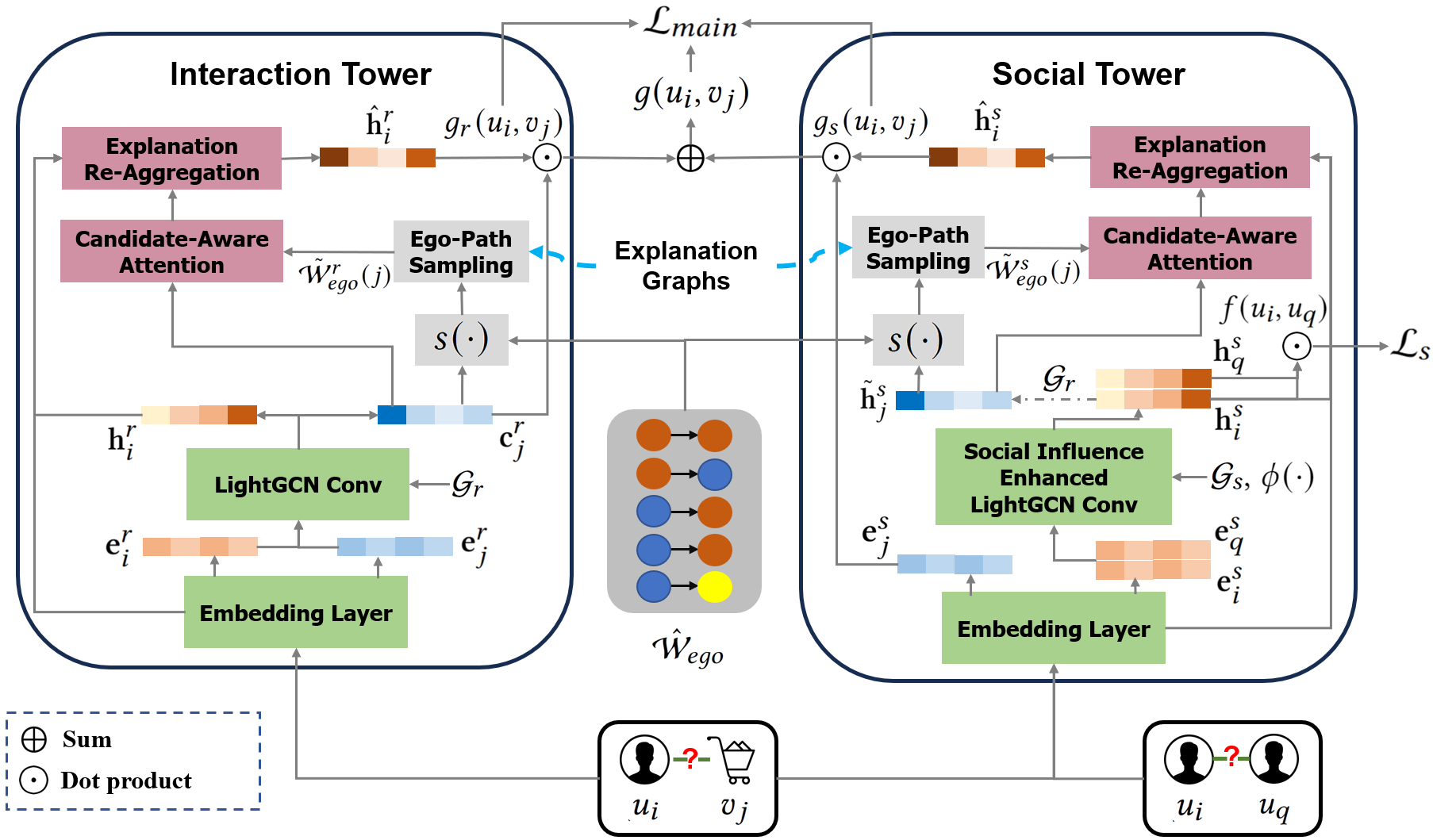}
    \caption{Overview of our proposed SoREX framework for self-explainable GNN-based social recommendation.}
    \label{fig:model}
\end{figure*}

The overview of our proposed SoREX is presented in Figure \ref{fig:model}. As shown, SoREX adopts a two-tower architecture, equipped with two components: \textit{ego-path sampling} and \textit{explanation re-aggregation}, which aim to generate explanations and explicitly relate explanations to predictions respectively. Specifically, the basic two-tower framework is composed of an interaction tower and a social influence-aware social tower, laying the foundation for factor-specific explanations. Given a disconnected user-item pair and the user's ego-path set, each tower independently performs user/item embedding, ego-path sampling and item ranking, with final ranking obtained by fusing both tower predictions. In each tower, we compute the path-level similarity between each ego-path and candidate item, which is regarded as the seed of independent Bernoulli sampling. The sampled ego-paths are treated as candidate-aware explanations for further candidate-wise comparison, and they are further re-aggregated into the output user representations based on candidate-aware attention, so that the relevant ego-paths can be explicitly related to the downstream ranking predictions. To further enhance the model's performance and improve the user embeddings' ability to capture genuine relationships between friends in social graphs, multi-task learning is employed. The auxiliary task of friend recommendation helps the user embeddings better learn true friend relationships, compensating for the noise in the social graph. The notations used throughout the method are summarized in Table~\ref{tb:notation}.

\subsection{Basic Two-Tower Framework}
\label{sec:base-framework}

To independently model social and user-item interaction factors and further lay the foundation for factor-specific explanation, we design a basic two-tower framework, which is composed of a social influence aware social tower and an interaction tower to learn from $\mathcal{G}_s$ and $\mathcal{G}_r$ respectively. Each tower is assigned with independent ID embeddings and GNN encoder, predicting its own ranking score.
We define the ID embedding matrix in social and interaction tower as $\mathbf{E}^s\in\mathbb{R}^{(m+n)\times d}$ and $\mathbf{E}^r \in \mathbb{R}^{(m+n)\times d}$ respectively, where $d$ is the embedding dimension. For interaction tower, we adopt $k_1$-layer LightGCN~\cite{lightgcn} to encode user/item representations for prediction:

\begin{align}
    & \mathbf{h}_i^r = \frac{1}{k_1 + 1}[\mathbf{h}_i^{r^{(0)}} + \sum\limits_{l=1}^{k_1}AGG^r(\mathbf{c}_j^{r^{(l)}} | j\in\mathcal{N}^r_i)], \\
    & \mathbf{c}_j^r = \frac{1}{k_1 + 1}[\mathbf{c}_j^{r^{(0)}} + \sum\limits_{l=1}^{k_1}AGG^r(\mathbf{h}_i^{r^{(l)}} | i\in\mathcal{N}^r_j)], \\
    & AGG^r(\mathbf{x}_i|i\in\mathcal{N}^r_q) = \sum\limits_{i\in\mathcal{N}^r_q} \frac{\mathbf{x}_i}{|\sqrt{\mathcal{N}^r_i}||\sqrt{\mathcal{N}^r_q}|},
\end{align}

\noindent where $\mathbf{h}^r$ and $\mathbf{c}^r$ are the encoded user and item embeddings, $\mathbf{h}_i^{r^{(0)}}$ and $\mathbf{c}_j^{r^{(0)}}$ are assigned with corresponding ID embeddings $\mathbf{e}_i^r\in\mathbb{R}^d$ and $\mathbf{e}_j^r\in\mathbb{R}^d$ respectively. $\mathcal{N}^r_q$ is the set of direct neighbors of $q$ on $\mathcal{G}_r$. The scoring function of interaction tower $g_r(u_i, v_j)$ is defined as the dot product of $\mathbf{h}_i^r$ and $\mathbf{c}_j^r$. 

As for the social tower, we adopt a mutant of LightGCN as the GNN encoder. Empirical studies in ~\cite{design} suggest that the influence of socially active users' purchase decisions on which of other users has no significant increase or decrease compared with inactive users. Thus, the neighborhood aggregation function in LightGCN is not suitable for user-item interaction modeling in social recommendation. 
As suggested in ~\cite{share}, socially connected users will have more influence on each other if they are of similar purchase behaviors. Therefore, we devise a social influence aware aggregation function and integrate it into original LightGCN architecture to encode $\mathcal{G}_s$. The derivation of encoded user embedding $\mathbf{h}^s$ is formulated as follows:

\begin{align}
    & \mathbf{h}_i^s = \frac{1}{k_2 + 1}[\mathbf{h}_i^{s^{(0)}} + \sum\limits_{l=1}^{k_2}AGG^s(\mathbf{h}_j^{s^{(l)}} | j\in\mathcal{N}^s_i)], \\
    & AGG^s(\mathbf{x}_i|i\in\mathcal{N}^s_q) = \sum\limits_{i\in\mathcal{N}^s_q}\alpha_i \mathbf{x}_i, \\
    & \alpha_i = \mathop{\text{softmax}}_{i\in \mathcal{N}^s_q} \phi(i, q),
\end{align}

\noindent where $k_2$ is the number of convolutional layers in social tower, $\mathbf{h}_i^{s^{(0)}}$ is assigned with ID embedding $\mathbf{e}_i^s\in\mathbb{R}^d$,  $\mathcal{N}^s_q$ is the set of direct neighbors of $q$ on $\mathcal{G}_s$. $\phi(i, q)$ is the social influence function, evaluating the strength of social connection between $u_i$ and $u_q$. We assume that a pair of socially connected users will have stronger social connections with each other if the overlap ratio of their interacted item sets is high. We could quantitatively measure such overlap with Jaccard similarity. However, we find that the difference of the Jaccard similarity among all available users is subtle. For simplicity, we adopt its square root to amplify the social influence difference. Therefore, $\phi(i,q)$ is formulated as follows:

\begin{equation}
    \phi(i, q) = \sqrt{\frac{|\mathcal{N}^r_i \cap \mathcal{N}^r_q|}{|\mathcal{N}^r_i \cup \mathcal{N}^r_q|}}.
\end{equation}

The scoring function of social tower $g_s(u_i, v_j)$ is defined as the dot product of $\mathbf{h}_i^s$ and the ID embedding of candidate item $\mathbf{e}_j^s$. With $g_s$ and $g_r$, the final scoring function $g$ of the basic framework is formulated as:

\begin{equation}
    g(u_i, v_j) = g_r(u_i, v_j) + g_s(u_i, v_j).
\end{equation}

Note that the basic framework is flexible for that the encoder in each tower could be replaced by any advanced GNN encoders. As for the potential impact, we leave it for future work.

\subsection{Ego-Path Based Explanation}
\label{sec:ego-path}

This section presents the explanation approach used in SoREX. We aim to provide substructure-based explanations to enhance transparency of social recommenders. 
Previous explainable recommendation methods for transparency enhancement fail to utilize pure structural information.

Although existing self-explainable GNNs are able to investigate complex explanatory substructures, they fail to provide comparative explanations for ranking predictions. 
To address these challenges, we propose to sample relevant subgraphs of $\mathcal{G}_{ego}$ for each candidate item $v_j$ and each concerned factor as candidate-aware and factor-specific explanation graph. This process is equivalent to sample subsets of $\mathcal{W}_{ego}$. Considering that the size of $\mathcal{W}_{ego}$ grows exponentially with the increase of hop number $k$, it is not feasible to perform path ranking over the whole $\mathcal{W}_{ego}$. Inspired by ~\cite{subgraph_gnn}, we randomly and uniformly sample $n_w$ walks originated from target user $u_i$ with maximum length $k$. Then we transform the sampled paths into the form of ego-paths introduced in section \ref{sec:preliminary}. This procedure essentially selects an ego-path subset $\hat{\mathcal{W}}_{ego}$ from all available ego-paths in $\mathcal{W}_{ego}$ to extract an even smaller explanation graph from. Note that repetitive walks will not be removed from $\hat{\mathcal{W}}_{ego}$ in order to implicitly reflect structural features of $\mathcal{G}_{ego}$.

Given candidate item $v_j$, the sampling probability for ego-path $w_t \in \hat{\mathcal{W}}_{ego}$ is determined by similarity between $w_t$ and $v_j$ in specific tower. We take the average similarity between candidate $v_j$ and each node $q$ in $w_t$ as the path-level similarity. We formulate $p_t^{r*}(v_j)$ and $p_t^{s*}(v_j)$, the sampling seed in interaction tower and social tower for $w_t$ in Eq. (\ref{eq:prob-inter}) and Eq. (\ref{eq:prob-soc}) respectively:

\begin{align}
    \label{eq:prob-inter}
    & p_t^{r*}(v_j) = \frac{1}{k-1}(\sum\limits_{q \in \mathcal{U} \cap w_t} s(\mathbf{h}_q^r, \mathbf{c}_j^r) + \sum\limits_{q \in \mathcal{V}\cap w_t}s(\mathbf{c}_q^r, \mathbf{c}_j^r)), \\
    \label{eq:prob-soc}
    & p_t^{s*}(v_j) = \frac{1}{k-1}(\sum\limits_{q \in \mathcal{U}\cap w_t} s(\mathbf{h}_q^s, \tilde{\mathbf{h}}_j^s) + \sum\limits_{q \in \mathcal{V}\cap w_t}s(\tilde{\mathbf{h}}_q^s, \tilde{\mathbf{h}}_j^s)), \\
    & \tilde{\mathbf{h}}_j^s = \left\{
    \begin{aligned}
        & \frac{1}{|\mathcal{N}^r_j|}\sum\limits_{i\in\mathcal{N}^r_j} \mathbf{h}_{i}^s, \quad|\mathcal{N}_j^r| > 0 \\
        & \mathbf{e}_j^s, \quad|\mathcal{N}_j^r| = 0
    \end{aligned}
    \right.
    ,
\end{align}

\noindent where $s(\cdot)$ is cosine similarity. The probability computation in interaction tower is directly based on the output representations of GNN encoder. In Eq. (\ref{eq:prob-inter}) and Eq. (\ref{eq:prob-soc}), the denominator is $k-1$ because we are excluding the target user when computing the path-level similarity. The value of 
$k-1$ corresponds to the length of the path excluding the starting node, as we are only interested in the relationships between the target user and other nodes in the path.
In social tower, the user representations used for probability computation are also the output of GNN encoder. Considering that item nodes do not participate in message propagation in $\mathcal{G}_s$, to fully leverage structural knowledge in social GNN encoder, we take the mean pooling of the GNN-encoded representations of users that have interacted with candidate $v_j$ as item representation used for probability computation, which corresponds to $\tilde{\mathbf{h}}_j^s$. 
If $v_j$ is a cold-start item with no interactions, we will replace $\tilde{\mathbf{h}}_j^s$ with $v_j$'s ID embedding. We further adjust $p_t^{r*}(v_j)$ and $p_t^{s*}(v_j)$ into range $[0, 1]$ as sampling probabilities $p_t^r(v_j)$ and $p_t^s(v_j)$:

\begin{equation}
    p_t^r(v_j) = \frac{p_t^{r*}(v_j) + 1}{2}, \quad p_t^s(v_j) = \frac{p_t^{s*}(v_j) + 1}{2}.
    \label{eq:path-sim}
\end{equation}

Then, we sample ego-path subsets $\tilde{\mathcal{W}}_{ego}^r(j)$ and $\tilde{\mathcal{W}}_{ego}^s(j)$ based on Bernoulli distribution $\beta_t^r \sim \text{Bern}(p_t^r(v_j))$ and $\beta_t^s \sim \text{Bern}(p_t^s(v_j))$ in interaction tower and social tower respectively. We only keep ego-paths with $\beta_t^* = 1$ in corresponding tower. To make sure the gradient w.r.t. $p_t^r(v_j)$ and $p_t^s(v_j)$ is computable, we apply gumble-softmax reparameterization trick~\cite{gumble}. To this end, we have obtained two ego-path subsets relevant to $v_j$ from the view of interaction and social factor respectively. They will be regarded as the factor-specific explanation graphs for $v_j$'s ranking result.

Our ego-path based explanations can effectively address aforementioned challenges. The sampled ego-paths can form dense explanation graphs. Ideally, ego-paths interweave with each other and form various motifs, enabling investigation of complex substructures based on their quantity and importance. Besides, 
we can easily compare different candidate-aware and factor-specific ego-path subsets and come up with comparative explanations from different perspectives, filling the gap of comparative explainability in GNN-based social recommenders.

\subsection{Explanation Re-Aggregation}
\label{sec:reaggr}

To make SoREX self-explainable, we need to relate explanation to predictions. Many self-explainable methods conduct downstream prediction directly based on explanations. To reduce information loss, we instead treat the sampled ego-path subsets as an emphasis of the relevant part of $\mathcal{G}_{ego}$ and perform explanation re-aggregation to relate the sampled ego-paths with final prediction. 

Explanation re-aggregation aims to aggregate information from sampled ego-paths into original GNN-encoded representations of target users in corresponding tower, such that the factor-specific and candidate-aware knowledge within $\mathcal{G}_{ego}$ can be emphasized. A hop-wise attention based node-level aggregation method is designed for this procedure. We take interaction tower as an example for illustration. Given GNN-encoded user embedding $\mathbf{h}_i^r$, candidate item $v_j$ and interaction-specific $v_j$-aware ego-path subset $\tilde{\mathcal{W}}_{ego}^r(j)$, we first compute attention in the fashion of Transformer~\cite{transformer} for each node $q$ in $w_t \in \tilde{\mathcal{W}}_{ego}^r(j)$ based on its cosine similarity with the GNN-encoded representation of $v_j$:

\begin{equation}
    a_{qj}^r = \left\{
    \begin{aligned}
        & \frac{s(\mathbf{h}_q^r, \mathbf{c}_j^r)}{\sqrt{d}}, q \in \mathcal{U} \\
        & \frac{s(\mathbf{c}_q^r, \mathbf{c}_j^r)}{\sqrt{d}}, q \in \mathcal{V}
    \end{aligned}
    \right.
    .
\end{equation}

We use $\tilde{\mathcal{W}}_{ego}^r(j)[l], l\in\{1, 2, ..., k\}$ denote the collection of the $l$-th hop node in $\forall w_t \in \tilde{\mathcal{W}}_{ego}^r(j)$. For node $q$ in the $l$-th hop, its normalized attention value $\alpha_{qj}^r$ is formulated as follows:

\begin{equation}
    \alpha_{qj}^r = \mathop{\text{softmax}}_{q \in \tilde{\mathcal{W}}_{ego}^r(j)[l]} a_{qj}^r .
\end{equation}

After hop-wise attention normalization, we perform ego-path aggregation via simple addition based on the ID embeddings of nodes involved in sampled ego-paths. The interaction-specific explanation enhanced final user embedding is defined as Eq. (\ref{eq:emphasis_final_emb}):

\begin{equation}
\label{eq:emphasis_final_emb}
    \hat{\mathbf{h}}_i^r = \frac{1}{k+1} (\mathbf{h}_i^r + \sum\limits_{w_t\in \tilde{\mathcal{W}}_{ego}^r(j)} \sum\limits_{q \in w_t} \alpha_{qj}^r \mathbf{e}_q^r).
\end{equation}

The attention weights $\alpha_{qj}^s$ and explanation enhanced final user embedding $\hat{\mathbf{h}}_i^s$ in social tower can be similarly defined with $\mathbf{h}_i^s, \tilde{\mathbf{h}}_j^s$ and $\mathbf{E}^s$. Given $\hat{\mathbf{h}}_i^s$ and $\hat{\mathbf{h}}_i^r$, we redefine $g_r(u_i, v_j)$ as the dot product of $\hat{\mathbf{h}}_i^r$ and $\mathbf{c}_j^r$, and redefine $g_s(u_i, v_j)$ as dot product of $\hat{\mathbf{h}}_i^s$ and $\mathbf{e}_j^s$. In Eq. (\ref{eq:emphasis_final_emb}), the denominator is $k+1$ because we include the target user in the aggregation process. This ensures that the representation of the user is influenced by both their original embedding and the embeddings of the nodes in the sampled ego-paths, including the starting node.

\subsection{Multi-Task Training}
\label{sec:objective}

We employ multi-task learning to optimize two tasks jointly in SoREX. The user-item interaction modeling task serves as the major task, and we adopt BPR loss for pairwise ranking. We conduct optimization for each tower's own predictions and the fused final predictions. The loss function $\mathcal{L}_{main}$ for the primary task is formulated as follows:

\begin{equation}
    \mathcal{L}_{main} = \sum\limits_{(u,v,v^-)\in \mathcal{E}_t} 
    -\text{log}\sigma(g_r(u, v) - g_r(u, v^-))
    -\text{log}\sigma(g_s(u, v) - g_s(u, v^-))
    -\text{log}\sigma(g(u, v) - g(u, v^-)),
\end{equation}

\noindent where $v$ and $v^-$ represent positive and negative sample respectively, and $\mathcal{E}_t$ is the training set. The other auxiliary task is friend recommendation task. The introduction of friend recommendation can help distill more supervision signals from social domain in the view of user-user relationships, instead of the relationships between social proximity and user-item interactions only. Recent work~\cite{cgcl} has demonstrated its positive effect for social recommendation. Therefore, we randomly sample positive and negative user pairs from $\mathcal{G}_s$, and conduct candidate user ranking and item ranking simultaneously. Considering that friend recommendation only leverages social networks, we only integrate this task into social tower. The loss function of friend recommendation $\mathcal{L}_s$ is defined as:

\begin{align}
    & f(u_i, u_q) = \mathbf{h}_i^s \cdot \mathbf{h}_q^{s^T}, \\
    & \mathcal{L}_s = \sum\limits_{(u, u^+, u^-)\in\mathcal{E}_s} -\text{log}\sigma(f(u, u^+) - f(u, u^-)),
\end{align}

\noindent where $f$ is the scoring function for user pairs. The overall objective function $\mathcal{L}$ is formulated as follows:

\begin{equation}
    \mathcal{L} = \mathcal{L}_{main} + \gamma \mathcal{L}_s + \lambda(\parallel \mathbf{E}^s\parallel_2 + \parallel\mathbf{E}^r\parallel_2),
\end{equation}

\noindent where the last term is the L2 regularization, and $\gamma, \lambda$ are hyperparameters to control the strength of regularization. Note that the only trainable parameters in SoREX are $\mathbf{E}^s$ and $\mathbf{E}^r$.

\subsection{Computational Complexity}

This subsection aims to analyze the time and memory complexity of SoREX. Considering that representative baseline methods (e.g. LightGCN~\cite{lightgcn}, Diffnet~\cite{diffnet} and DESIGN~\cite{design}) essentially adopt different graph convolution strategies over the same joint graph $\mathcal{G}$, we compare our SoREX with the more computationally efficient method LightGCN and the multi-tower alike method DESIGN. The time and memory complexity of SoREX and the selected baselines are listed in Table \ref{tab_cc}. Suppose that $|\mathcal{U}|$ and $|\mathcal{V}|$ are the size of user set and item set, while $|\mathcal{E}|$ is the size of edge set of $\mathcal{G}$. Given the dimension of learnable embeddings $d$ and the layer number $L$, the time complexity of LightGCN and SoREX is $O(L|\mathcal{E}|d)$ and $O(L|\mathcal{E}|d + n_w|\mathcal{E}_r|d)$ respectively. DESIGN essentially performs message propagation over $\mathcal{G}$ for twice, so the time complexity of DESIGN is also $O(L|\mathcal{E}|d)$. The optimal $n_w$ usually satisfies $L \ll n_w \ll |\mathcal{E}|$. Therefore, the time complexity of SoREX is one to two orders of magnitude greater than that of LightGCN and DESIGN, which is acceptable when dealing with sparse networks. On the other hand, SoREX has no other trainable parameters except for the ID embeddings, which is same as LightGCN and DESIGN. Considering that the ego-path sampling procedures are real-time, the memory costs of SoREX is nearly the twice of LightGCN. Therefore, the memory complexity of SoREX and LightGCN are both $O(L(|\mathcal{U}|+|\mathcal{V}|)d)$. Items are not assigned with ID embeddings in DESIGN, so the memory complexity of DESIGN is $O(L|\mathcal{U}|d)$. In summary, although the time and memory complexity of SoREX are not superior, we aim to achieve the necessary trade-off between efficiency and explainability, while minimizing the unnecessary computational costs.

\label{sec:complexity}

\begin{table}[htbp]
  \centering
  \caption{Computational Complexity Comparison\textcolor{blue}{.}}
    \begin{tabular}{ccc}
    \hline
     & Time & Memory \\
    \hline
    LightGCN & $O(L|\mathcal{E}|d)$ & $O(L(|\mathcal{U}|+
    |\mathcal{V}|)d)$\\
    DESIGN & $O(L|\mathcal{E}|d)$ & $O(L|\mathcal{U}|d)$ \\
    SoREX & $O(L|\mathcal{E}|d + n_w|\mathcal{E}_r|d)$ & $O(L(|\mathcal{U}|+ |\mathcal{V}|)d)$ \\
    \hline
    \end{tabular}%
  \label{tab_cc}%
\end{table}%

\section{Experiments}
\subsection{Experimental Setup}

 \textit{Datasets.} We evaluate our SoREX on several widely used benchmark datasets, including three sparse datasets \textit{Yelp\footnote{https://github.com/librahu/HIN-Datasets-for-Recommendation-and-Network-Embedding}, Flickr\footnote{https://www.flickr.com/}}, \textit{Ciao}\footnote{https://www.cse.msu.edu/tangjili/datasetcode/truststudy.htm} and a dense dataset \textit{LastFM}\footnote{https://files.grouplens.org/datasets/hetrec2011}. Yelp is an online location-based social network. Flickr is an online image-based social sharing platform with a whom-trust-whom social network. Ciao is a popular social networking website. LastFM consists of a user-artist interaction network and a user friendship network. Consistent with previous work~\cite{diffnet}, we remove users and items with less than two interaction records. For datasets with ratings like Ciao, we only keep links with ratings no less than 4. Considering that all the selected datasets have no temporal information, we randomly split each dataset into train/validation/test sets at a ratio of 80\%/10\%/10\%. Detailed preprocessed dataset statistics are presented in Table \ref{tab:dataset}.

\begin{table}[htbp]
  \centering
  \caption{Preprocessed Dataset Statistics.}
    \begin{tabular}{c|ccccc}
    \hline
    Dataset & \#User & \#Item & \#Interaction & Density & \#Social \\
    \hline 
    Yelp  & 17,220 & 35,351 & 205,529 & 0.034\% & 143,609 \\
    Flickr & 8,137  & 76,190 & 320,775 &0.052\% & 182,078 \\
    Ciao  & 6,788  & 77,248 & 206,143 & 0.039\% & 110,383 \\
    LastFM & 1,880 & 3,933 & 75,228 & 1.017\% & 25,260 \\
    \hline
    \end{tabular}%
  \label{tab:dataset}%
\end{table}%

 \textit{Baseline Methods.} To validate the effectiveness of SoREX, we select two groups of baselines for comparison: 
 
 \noindent(1) GNN-based social recommendation baselines:
 
\begin{itemize}
    \item \textit{LightGCN}~\cite{lightgcn} is a popular GNN model for general recommendation task. It is characterized by recursive graph convolution without linear transformation and non-linear activation function.
    \item \textit{LightGCN$^S$} is the base model of CGCL~\cite{cgcl}, which is a friend recommendation enhanced LightGCN, conducting additional message propagation on social network to predict social links.
    \item \textit{SocialLGN}~\cite{sociallgn} performs message propagation on both $\mathcal{G}_r$ and $\mathcal{G}_s$ in each GNN layer to fuse their knowledge.
    \item \textit{Diffnet}~\cite{diffnet} is a layer-wise social influence propagation model proposed to simulate the social influence diffusion process.
    \item \textit{Diffnet++}~\cite{diffnet++} is an extension of Diffnet, modeling both social influence diffusion and user preference diffusion in a unified framework.
    \item \textit{MHCN}~\cite{mhcn} is a multi-channel hypergraph convolutional network. We remove the self-supervised learning related components in MHCN and test the effectiveness of its proposed model architecture.
    \item \textit{DESIGN}~\cite{design} leverages multiple GNN encoders to learn from different graphs and fuse their knowledge via knowledge distillation. 
    \item \textit{GBSR}~\cite{gbsr} leverages graph information bottleneck theory to identify and remove redundant social links, so that minimal but sufficient social network could be used for enhanced social recommendation. 
\end{itemize}

\noindent(2) Self-explainable GNN or recommendation baselines:

\begin{itemize}
    \item \textit{ConPI}~\cite{conpi} compares the similarity between neighbor sets of node pairs for both link prediction and explanation. Two versions of ConPI are devised. ConPI-Node identifies similar neighbors as explanations, while ConPI-Pair leverages similar neighbor pairs. We only test ConPI-Node here for the excessive memory complexity of ConPI-Pair.
    \item \textit{ECF}~\cite{ecf} is a self-explainable collaborative filter method, which aims to discover taste clusters from user-item interactions. The approximation relationships between learned taste clusters and items are used for explanation. Item tags are not adopted for cluster description in our experiment, while only user-item interaction links are used in the evaluation of ECF.
    \item \textit{GSAT}~\cite{gsat} is a self-explainable attention-based GNN proposed to inject learnable randomness into GNN with information bottleneck theory and sample dense explanatory subgraphs without spurious correlations.
    \item \textit{PxGNN}~\cite{pxgnn} is a prototype-based self-explainable GNN, which aims to generate prototype graphs for both prediction and prototype-based explanations. To transplant PxGNN into link prediction task, we set the amount of node class as 1, so that all the nodes share the same set of prototype graphs.
\end{itemize}
 
Note that both ConPI and GSAT are trained on the joint graph $\mathcal{G}$.
We do not adopt other representative self-explainable GNN baselines like SE-GNN~\cite{se-gnn}, ProtGNN~\cite{protgnn}, PGIB~\cite{lri} and DisC~\cite{disc} for comparison, because they are specifically designed for classification tasks. It is infeasible to extract node pair aware explanatory subgraphs via these methods, and thus it is non-trivial to directly transfer them to our task. We also exclude social recommendation baselines SEPT~\cite{sept}, DcRec~\cite{derec} and CGCL~\cite{cgcl}, which focus on self-supervised learning (SSL) for social recommenders. For recent social network denoising based methods~\cite{gbsr,recdiff,gdssl}, we only select GBSR for evaluation because their underlying ideas are similar. We would like to compare the superiority of different methods from the perspective of model architecture, so we exclude the SSL related components in the three baselines and test their underlying models, which correspond to LightGCN and LightGCN$^S$. 

\textit{Evaluation Metrics.} Following ~\cite{design}, we adopt Hit Rate (HR@$K$) and Normalized Discounted Cumulative Gain (NDCG@$K$) as evaluation metrics. HR@$K$ evaluates the accuracy of recommendation, while NDCG@$K$ can reflect the quality of item ranking. We set $K=10$ for both HR@$K$ and NDCG@$K$.

\textit{Implementation Details.}
We implement our framework and all selected baselines with PyTorch-Geometric~\cite{pyg_lib}. We fix the ID embedding dimension size $d$ as 64 for all methods. For SoREX, hop number of ego-paths $k$, layer number of GNN encoders $k_1$ and $k_2$ are all tuned from $\{1,2,3\}$. $n_w$, the size of $\hat{\mathcal{W}}_{ego}$, is tuned from $\{50, 100, 150, 200, 300, 500\}$. The friend recommendation task coefficient $\gamma$ is tuned from $\{0.1,0.2,...,1.0\}$. For LightGCN and LightGCN$^S$, we adopt 2-layer GNN architecture. The coefficient of auxiliary task in LightGCN$^S$ is tuned from the same range as $\gamma$. For all other baselines, we adopt the recommended settings in their paper. Adam optimizer~\cite{adam} is adopted for the optimization of all models with learning rate 0.001. The batch size is tuned from $\{256,512,1024\}$, and L2 regularization coefficient $\lambda$ is set as 0.001. Following ~\cite{diffnet}, for all methods, we randomly sample 10 and 1000 negative samples for each training and validating sample respectively. During testing, we rank all items that target users have not interacted with. Our code and data are available at https://github.com/antman9914/SoREX.

\begin{table*}[htbp]
  \centering
  \caption{Overall performance. Boldfaced and underlined scores are the best and the second-best ones respectively. The improvements achieved by SoREX are statistically significant ($p$-value $\ll$ 0.05).}
    \begin{tabular}{c|p{1.25cm}<{\centering}|p{1.25cm}<{\centering}|p{1.25cm}<{\centering}|p{1.25cm}<{\centering}|p{1.25cm}<{\centering}|p{1.25cm}<{\centering}|p{1.25cm}<{\centering}|p{1.25cm}<{\centering}}
    \hline
          & \multicolumn{2}{c|}{Yelp} & \multicolumn{2}{c|}{Flickr} & \multicolumn{2}{c|}{Ciao} & \multicolumn{2}{c}{LastFM} \\
\cline{2-9}          & HR@10 & NDCG  & HR@10  & NDCG  & HR@10 &  NDCG & HR@10 & NDCG \\
    \hline
    LightGCN & 0.0192 & 0.0089 & \underline{0.0045} & \underline{0.0023} & 0.0377 &  0.0205 & 0.1875  & 0.1042 \\
    LightGCN$^S$ & \underline{0.0215}  & \underline{0.0100} & 0.0043  & 0.0021 & \underline{0.0384} & \underline{0.0212} & 0.1884 & \underline{0.1064} \\
    SocialLGN & 0.0179 &  0.0086 & 0.0040 & 0.0020 & 0.0355 & 0.0180 & 0.1650  & 0.0941 \\
    Diffnet & 0.0187 & 0.0083 & 0.0026  & 0.0012 & 0.0224 & 0.0096 & 0.1497 & 0.0822  \\
    Diffnet++ & 0.0155 & 0.0076 & 0.0031 & 0.0014 & 0.0267 & 0.0132 & 0.1767 & 0.0987 \\
    MHCN  & 0.0183 & 0.0094 & 0.0044 & 0.0021 & 0.0326 & 0.0162 & \underline{0.1931} & 0.1029 \\
    GBSR & 0.0206 & 0.0097  & 0.0042 & 0.0021 & 0.0363 & 0.0182 & 0.1906 & 0.1045
    \\
    DESIGN & 0.0199  & 0.0091 & 0.0038 & 0.0019 & 0.0344 & 0.0187 & 0.1577 & 0.0904 \\
    \hline
    ConPI & 0.0053 & 0.0024 & 0.0010  & 0.0005 & 0.0051 & 0.0025  & 0.0691 & 0.0294 \\
    ECF & 0.0174 & 0.0082 & 0.0038 & 0.0017 & 0.0345 & 0.0177  & 0.1635 & 0.0952 \\
    GSAT  & 0.0197 & 0.0087 & 0.0040  & 0.0021 & 0.0324  & 0.0167 & 0.1810 & 0.1003 \\
    PxGNN & 0.0185 & 0.0080 & 0.0034 & 0.0015 & 0.0308 & 0.0149 & 0.1664 & 0.0985 \\
    \hline
    SoREX  & \textbf{0.0227} & \textbf{0.0104} &\textbf{0.0047}  & \textbf{0.0024} & \textbf{0.0402} & \textbf{0.0221}  & \textbf{0.1998} & \textbf{0.1144} \\
    \hline
    Improvement & 5.58\% & 4.00\% & 4.44\% & 4.35\% & 4.69\% &  4.25\%  & 3.47\% & 7.52\% \\
    \hline
    \end{tabular}%
  \label{tab:result}%
\end{table*}%

\subsection{Main Results}

We report the average evaluation results of 5 runs with different random seeds in Table \ref{tab:result}, where NDCG is short for NDCG@10 results.
Due to the existence of randomness in SoREX and GSAT, 
we take the average performance of 5 times of tests for each run. The relative improvement is also presented. We have the following observations: (1) Our SoREX significantly outperforms other baseline methods on all metrics and all datasets, which demonstrates the superiority of our proposed model architecture. (2) LightGCN and LightGCN$^S$ are the most competitive baselines, which demonstrates the superiority of pure message propagation structure in social recommendation. (3) The performance of baselines specifically designed for social recommendation significantly depends on dataset. (4) The performance of ConPI, GSAT and PxGNN are inferior in social recommendation, which could be the result of inappropriate underlying model architecture. 

Compared with the selected baselines, we attribute the success of SoREX to following designs: (1) an independent two-tower architecture, which enables independent modeling of both social and user-item interaction factors. (2) explanation re-aggregation, which emphasizes the relevant subgraphs of target users' ego-nets and assists the final predictions. (3) The integration of friend recommendation task, whose effectiveness is also verified in LightGCN$^S$.

\subsection{Hyperparameter Analysis}

To enable a more in-depth empirical analysis, we conduct a hyperparameter study focusing on three key aspects: (1) the multi-task balancing parameter $\gamma$, (2) the number of layers $k_1$ and $k_2$ in the social graph and the recommendation graph, respectively, and (3) the choice of the social aggregation function.

\begin{figure}[!t]
\centering
\subfigure[]{
    \hspace{-25pt}
        \includegraphics[width=0.22\linewidth]{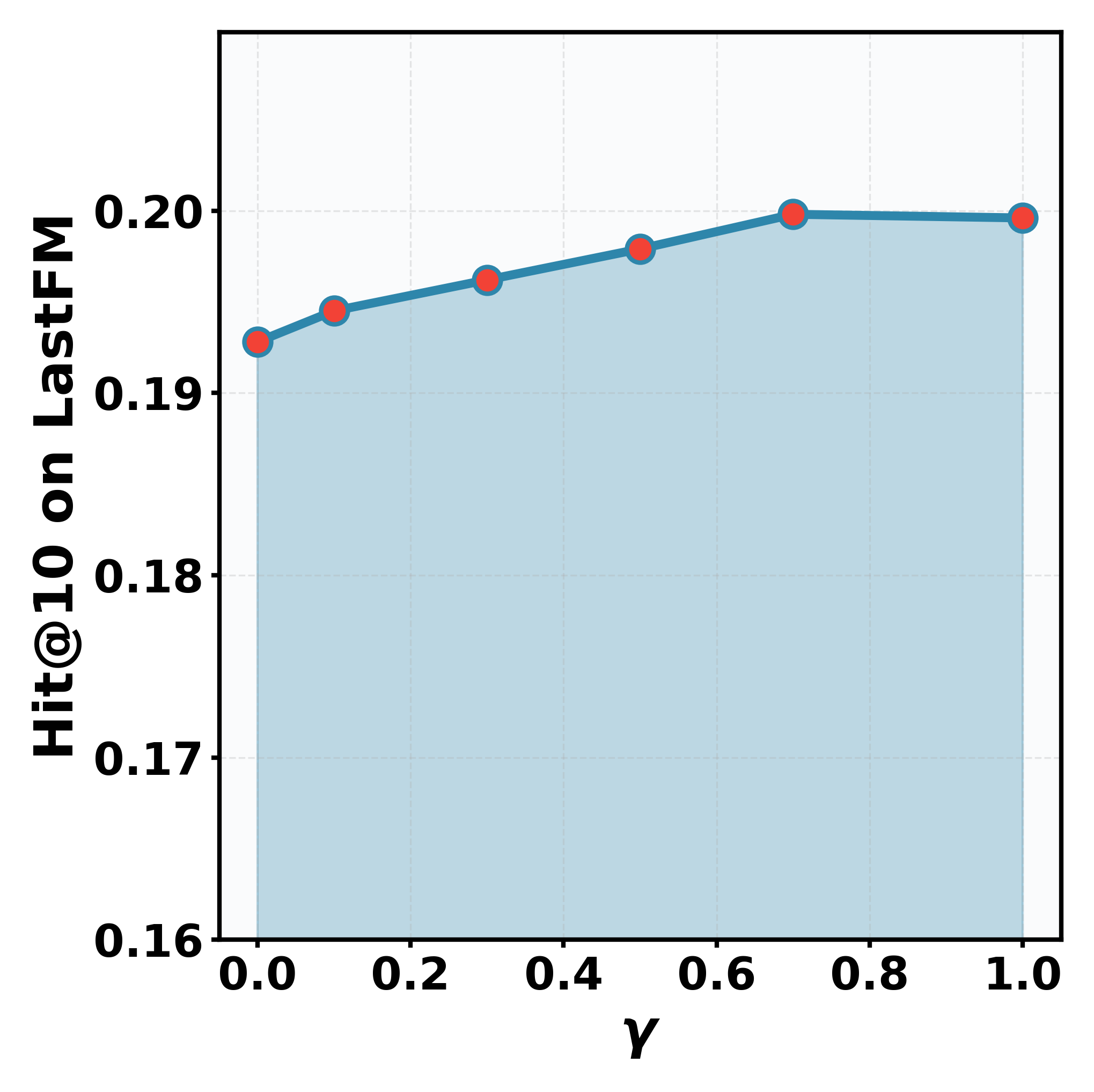}
        \label{fig:results_4_1}
    }
    \hspace{25pt}
    \subfigure[]{
    \hspace{-25pt}
        \includegraphics[width=0.22\linewidth]{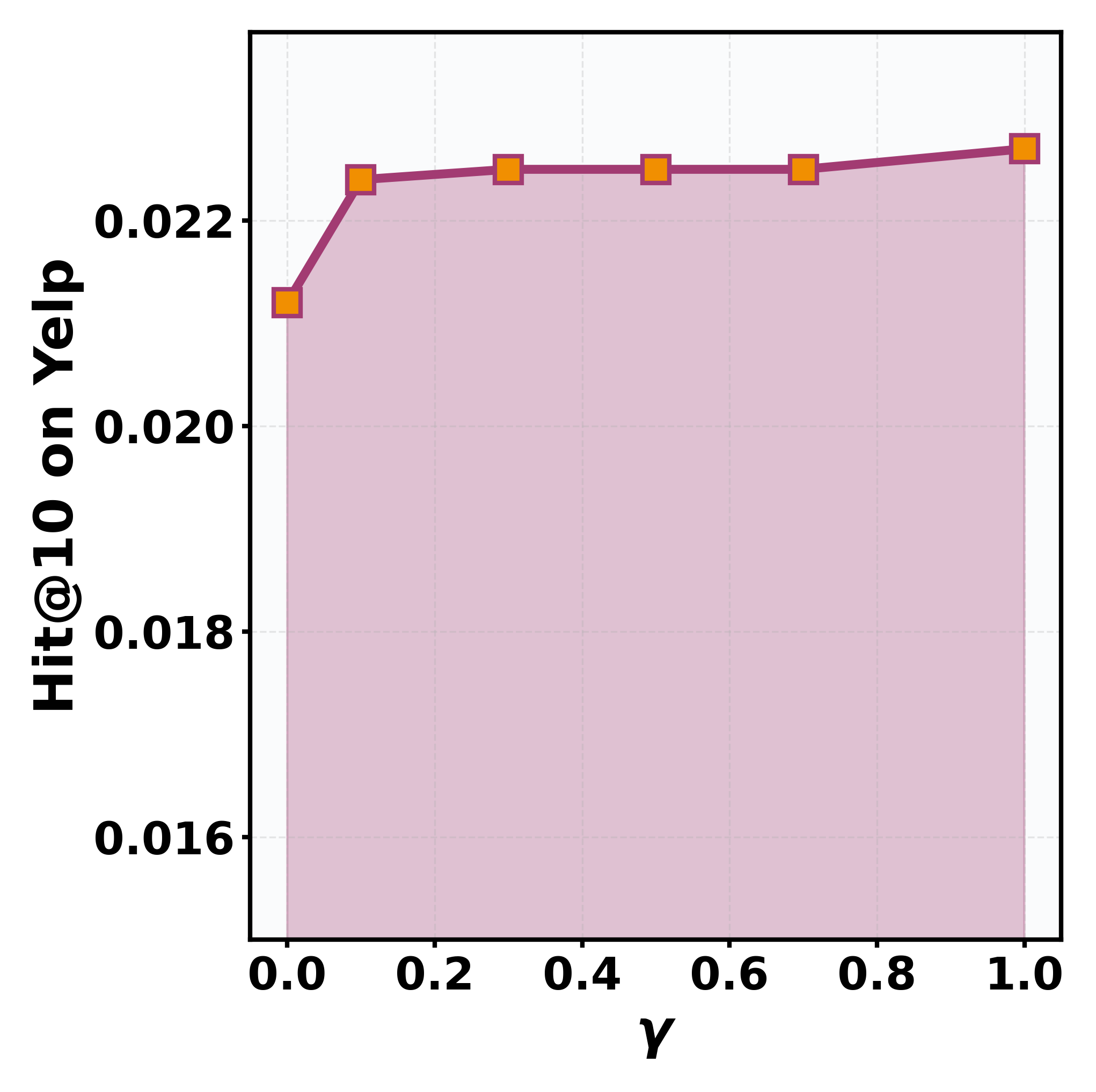}
        \label{fig:results_4_2}
    }
    \hspace{25pt}
    \subfigure[]{
    \hspace{-25pt}
        \includegraphics[width=0.22\linewidth]{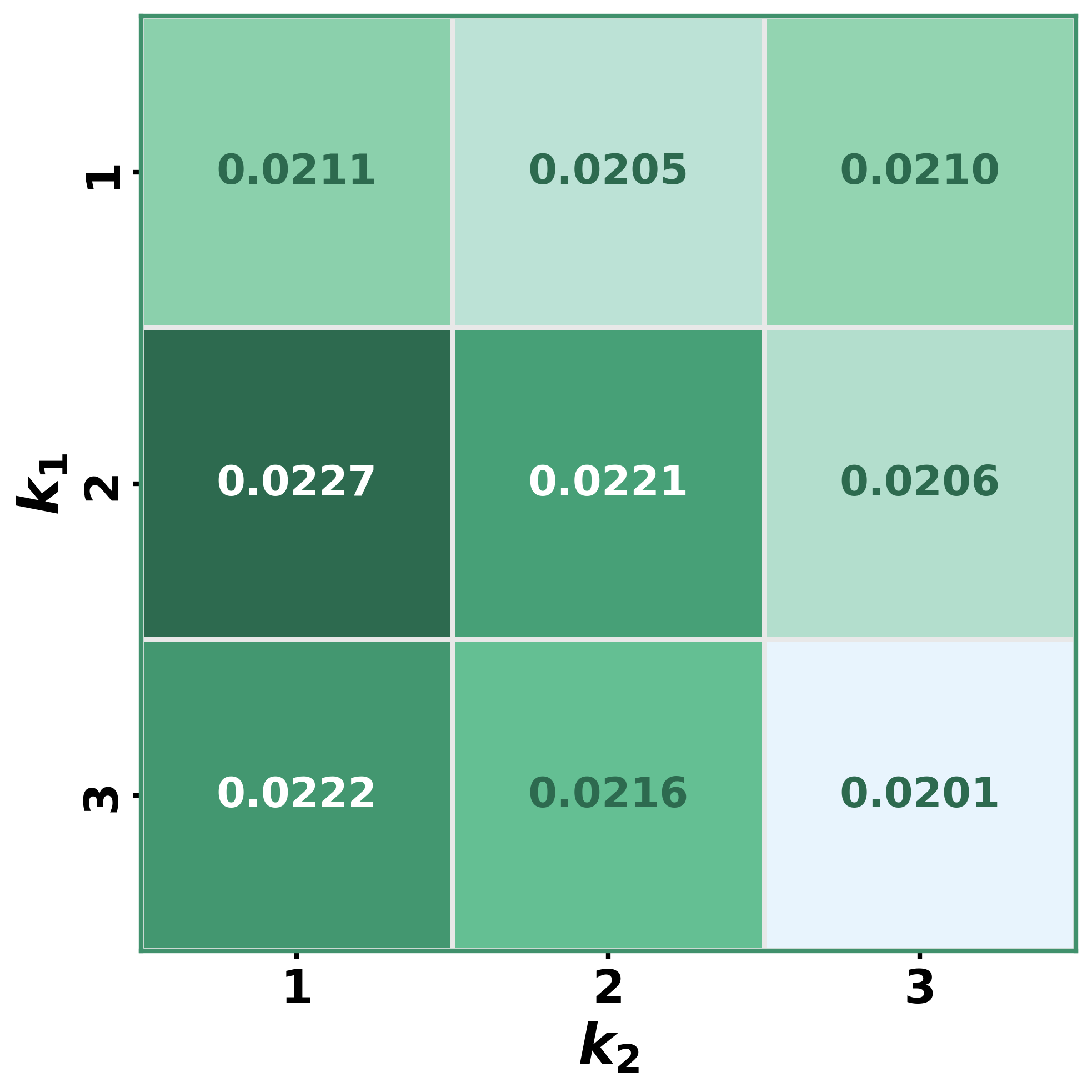}
        \label{fig:results_4_3}
    }
    \hspace{25pt}
    \subfigure[]{
    \hspace{-25pt}
        \includegraphics[width=0.22\linewidth]{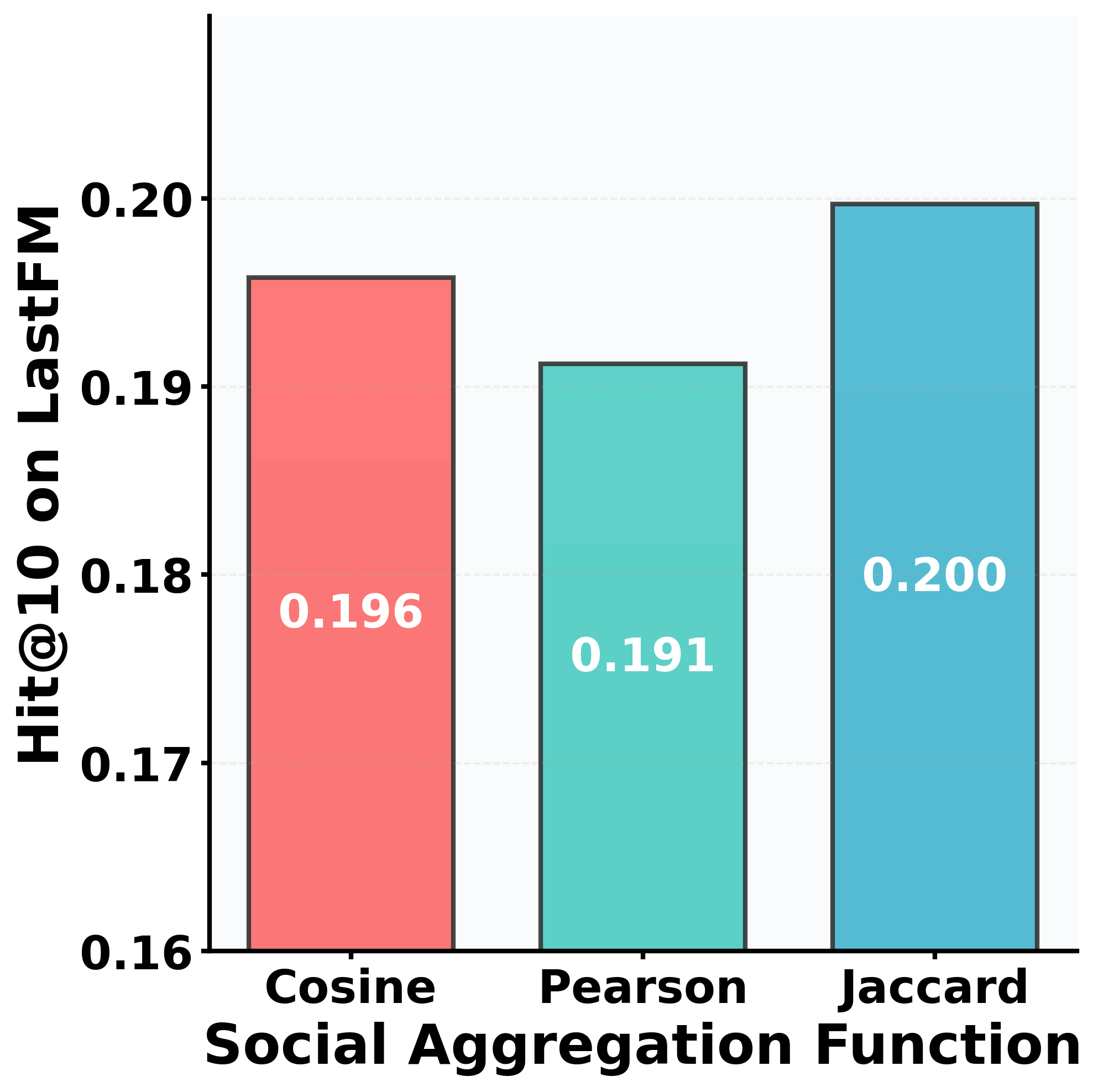}
        \label{fig:results_4_4}
    }
\caption{Hyperparameter sensitivity analysis. The figure presents the effects of varying the key hyperparameters on the model's performance, including the choice of the aggregation function, the multi-task balancing parameter $\gamma$, and the number of layers $k_1$ and $k_2$ in both graphs. The rightmost bar (\emph{Jaccard}) is significantly better than Cosine and Pearson at the 5\% level (paired \textit{t}-test, $p<0.05$).}
\label{fig:hyperparameter_analysis}
\end{figure}

As shown in Figures~\ref{fig:results_4_1} and~\ref{fig:results_4_2}, the model performs relatively poorly when $\gamma$ is set to 0. Performance stabilizes when $\gamma$ ranges between 0.1 and 1.0, underscoring the importance of the multi-task design in enhancing social recommendation. Figure~\ref{fig:results_4_3} illustrates the effect of $k_1$ and $k_2$ on Hit@10 performance using the Yelp dataset. Notably, the social GNN ($k_2$) achieves better results with fewer layers (1–2), likely due to the higher noise level in social graphs compared to user–item interaction graphs.

In our framework, social influence is modeled using Jaccard similarity, which captures the proportion of shared neighbors between users. To assess the effectiveness of this design, we compare Jaccard with two commonly used alternatives: cosine similarity and the Pearson correlation coefficient. As shown in Fig.~\ref{fig:results_4_4}, the Jaccard-based strategy consistently outperforms both. This superior performance stems from Jaccard's ability to directly reflect discrete overlap in social connections—an especially meaningful signal in social networks, where shared neighbors often imply stronger influence and homophily. In contrast, cosine similarity relies on the orientation of user embedding vectors, making it more susceptible to noise, while Pearson correlation is tailored for continuous variables and performs poorly with binary relational data such as friendship links (1 or 0). These results confirm that Jaccard similarity is more suitable for modeling social influence in our setting.

\subsection{Ablation Study}

To demonstrate the effectiveness of our proposed components, we conduct two series of ablation experiments on Yelp dataset: \textit{(1) General ablation studies:} We first remove auxiliary objective function $\mathcal{L}_s$ (w/o $\mathcal{L}_s$), our proposed explanation re-aggregation procedure (w/o re-aggr) and all components related to social tower (w/o soc tower) respectively. Then, we replace the social influence aware LightGCN encoder in social tower with the original LightGCN (w/o soc impact). \textit{(2) Alternative design studies:} We first attempt to replace the ego-path sampling procedure with top-$K$ ego-path extraction to generate sparse explanation graph (w/ top-$K$ ego-path). Then, referring to ~\cite{gsat}, we test the effect of additional information bottleneck (IB) based constraints on the sampling probability distributions in both towers, trying to remove potential spurious correlations in explanatory subgraphs (w/ IB constraint). We also try to share the same ID embeddings between the two towers (w/ shared ID), and replace user-transformed item representations $\tilde{h}_j^s$ used for sampling probability computation in social tower with their corresponding ID embeddings $e_j^s$ (w/o trans item emb).

\begin{table}[htbp]
  \centering
  \caption{Ablation study on Yelp.}
    \begin{tabular}{l|p{1.2cm}<{\centering}|p{1.2cm}<{\centering}|p{1.2cm}<{\centering}}
    \hline
               & HR@10  & MRR   & NDCG \\
    \hline
     SoREX  & \textbf{0.0227} & \textbf{0.0113} & \textbf{0.0104} \\
          w/o $\mathcal{L}_s$ & 0.0207 & 0.0103 & 0.0091 \\
          w/o re-aggr & 0.0215 & 0.0109 & 0.0100 \\
          w/o soc impact & 0.0209 & 0.0106 & 0.0095 \\
          w/o soc tower & 0.0205 & 0.0101 & 0.0092 \\
          \hline
          w/ top-$K$ ego-path & 0.0200  &  0.0100 & 0.0091  \\
          w/ IB constraint & 0.0217 & 0.0110 & 0.0100 \\
          w/ shared ID & 0.0209 & 0.0109 & 0.0098 \\
          w/o trans item emb &  0.0210 & 0.0106  & 0.0095  \\ 
    \hline
    \end{tabular}%
  \label{tab:ablation}%
\end{table}%

Based on the results in Table \ref{tab:ablation}, we can find that the full version of SoREX outperforms all other variants. We have following observations: 
\textit{For verification of the proposed components:} 
(1) Both social context and friend recommendation auxiliary task are significantly beneficial for recommendation. Their removal both result in performance drop of approximately 10\%. 
(2) The removal of social influence modeling causes significant performance drop, which demonstrates the effectiveness of user preference based social influence. 
(3) Utilizing information from explanation graphs can improve predictive accuracy by 4-5\%, indicating the necessity for emphasizing relevant neighborhood of target user.
\textit{For alternative designs:} 
(1)The variant using sparse explanation graphs is inferior than the full SoREX, because dense graphs can keep ego-paths that are less relevant but important for item ranking. 
(2) Integration of IB based constraint does not improve accuracy, indicating that the constraint is not suitable for our proposed attention mechanism. 
(3) Based on the performance drop in variant "w/ shared ID", independent modeling of social and interaction information is necessary. 
(4) User-transformed item representation is helpful for fully utilization of structural knowledge within social GNN encoder.

\subsection{Stability Analysis}

Ego-path generation in SoREX relies on random walk sampling, which introduces inherent randomness that may affect prediction stability—particularly in dense networks or for high-degree nodes. To examine this effect, we conduct experiments on two representative datasets: LastFM (a relatively dense dataset with density $>$1\%) and Yelp (the sparsest among the four, with a density of 0.034\%). Comparing these datasets allows us to evaluate SoREX under different levels of sparsity: in sparse graphs like Yelp, lower node degrees and smaller path selection spaces naturally reduce sampling noise, leading to more stable performance. This dataset-specific analysis provides deeper insight into SoREX’s robustness across different scenarios.

\begin{figure}
    \centering
    \subfigure[]{
    \hspace{-25pt}
        \includegraphics[width=0.22\linewidth]{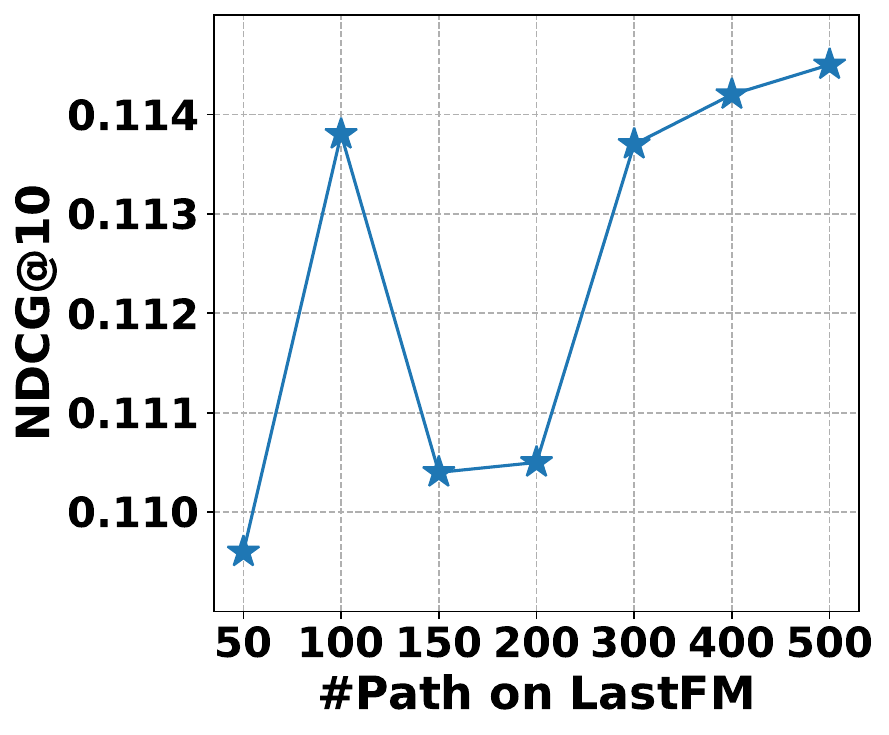}
        \label{fig:stability_4_1}
    }
    \hspace{25pt}
    \subfigure[]{
    \hspace{-25pt}
        \includegraphics[width=0.22\linewidth]{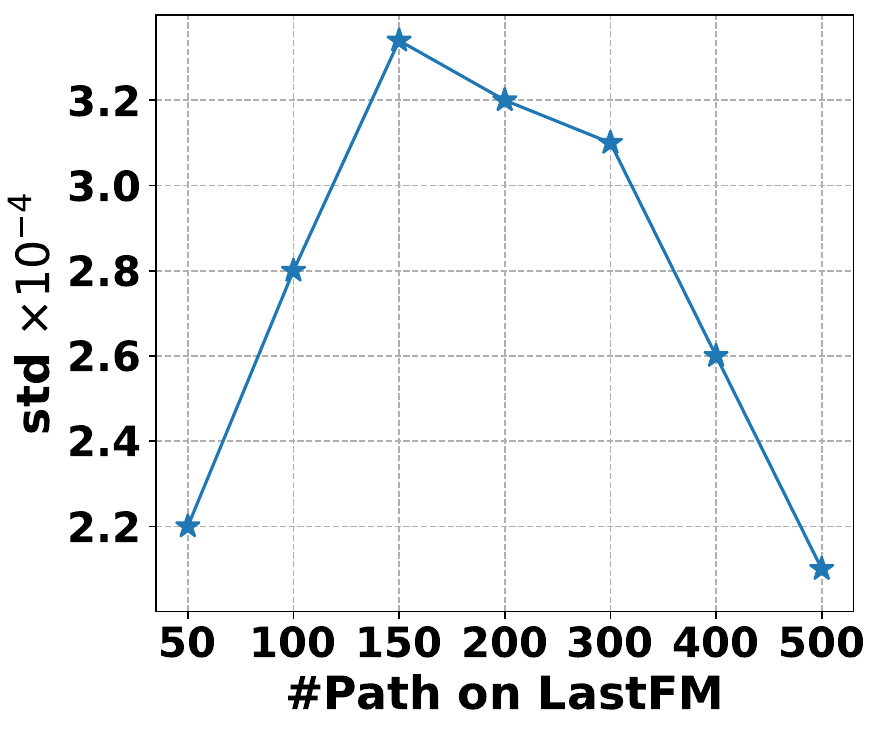}
        \label{fig:stability_4_2}
    }
    \hspace{25pt}
    \subfigure[]{
    \hspace{-25pt}
        \includegraphics[width=0.22\linewidth]{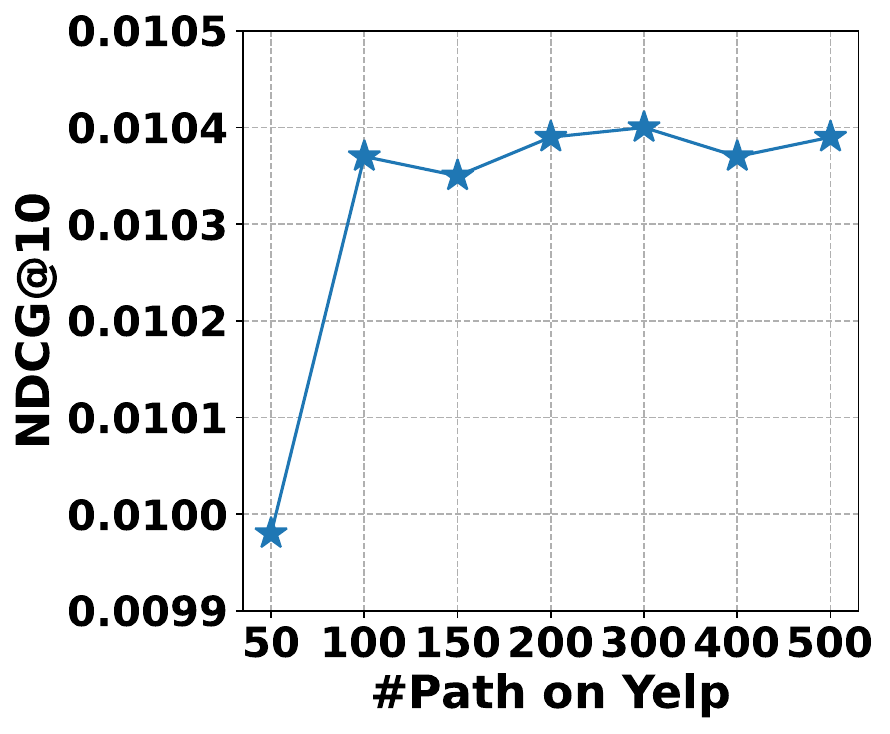}
        \label{fig:stability_4_3}
    }
    \hspace{25pt}
    \subfigure[]{
    \hspace{-25pt}
        \includegraphics[width=0.22\linewidth]{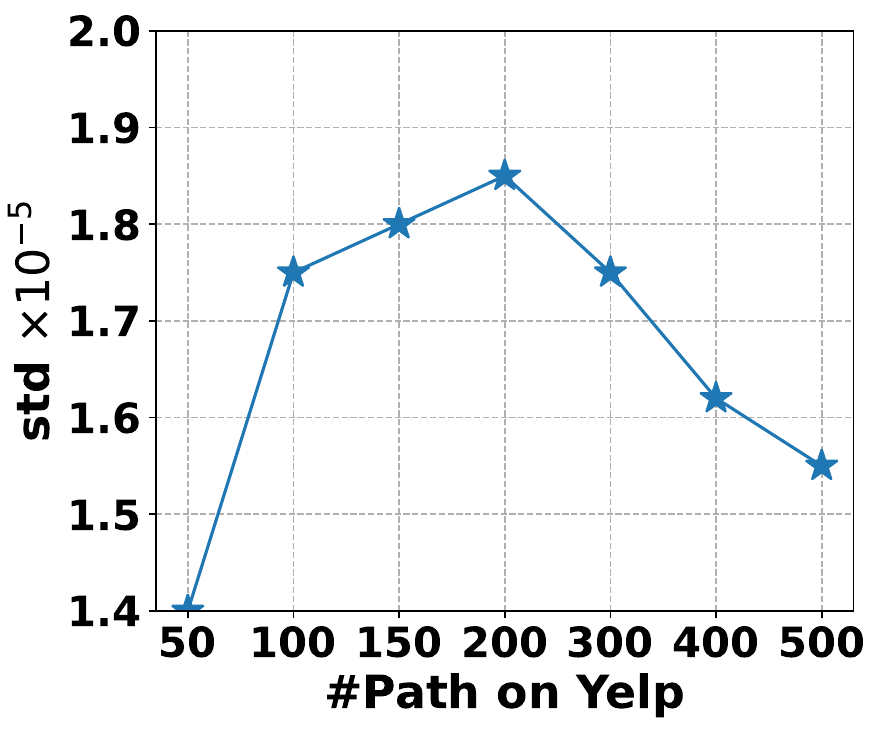}
        \label{fig:stability_4_4}
    }
    \caption{Stability analysis of SoREX with respect to the number of ego-paths on LastFM and Yelp. (a) and (b) report the average and standard deviation of NDCG@10 on LastFM as the number of ego-paths varies. (c) and (d) show the corresponding results for Yelp.}
    \label{fig:stability}
\end{figure}

In our experiments, we fix the random seed and retrain SoREX multiple times, varying the number of sampled ego-paths. For each path quantity, we perform 5 test runs and report the average NDCG@10 and standard deviation in Figure~\ref{fig:stability} to analyze sensitivity.
Our observations are as follows:
(1) Incorporating ego-paths improves recommendation performance, yet simply increasing their number does not always yield further gains, as shown in Figure~\ref{fig:stability_4_1} and Figure~\ref{fig:stability_4_3}. (2) While incorporating more ego-paths generally improves stability, this holds true only after reaching a minimum threshold. Below this threshold, adding paths may increase variance, likely due to insufficient sampling to represent the ego-network's statistical properties. (3) Despite minor fluctuations in standard deviation with varying ego-path quantities, the performance oscillation on LastFM remains within a narrow range of 0.2–0.3\%(Figure~\ref{fig:stability_4_2}), indicating high robustness. (4) As shown in Figure~\ref{fig:stability_4_4}, SoREX achieves even more stable results on the sparser Yelp dataset. The smaller node degrees and reduced path selection space lower the potential noise, allowing the model to reach optimal performance with fewer paths.

\subsection{Explanation Analysis}

We analyze our generated explanations from both qualitative and quantitative aspects. We first conduct case study for instance-level qualitative analysis to further illustrate the idea of comparative explanation, and then provide systematic pattern analysis to demonstrate SoREX's ability to investigate important substructures for recommendation. We also conduct quantitative analysis based on fidelity score~\cite{fidelity}. All discussions are based on the setting $k=2$. Only testing samples with ground truth items ranked within top-5 are considered.

\begin{figure}
    \centering
    \subfigure[]{
        \includegraphics[width=0.4\linewidth]{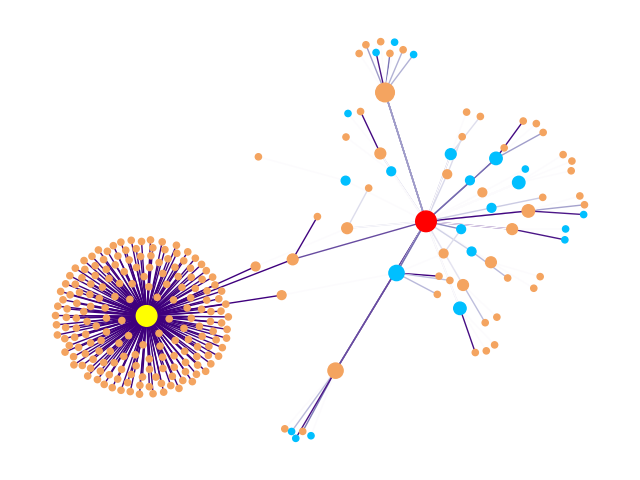}
        \label{fig:pos-co-case}
    }
    \subfigure[]{
        \includegraphics[width=0.4\linewidth]{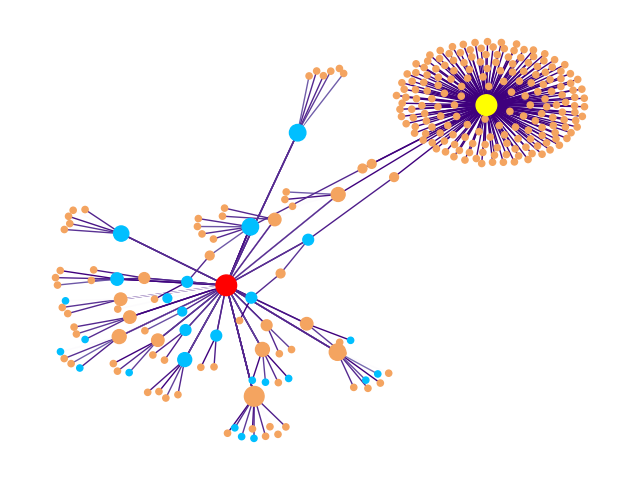}
        \label{fig:pos-soc-case}
    }
    \\
    \subfigure[]{
        \includegraphics[width=0.4\linewidth]{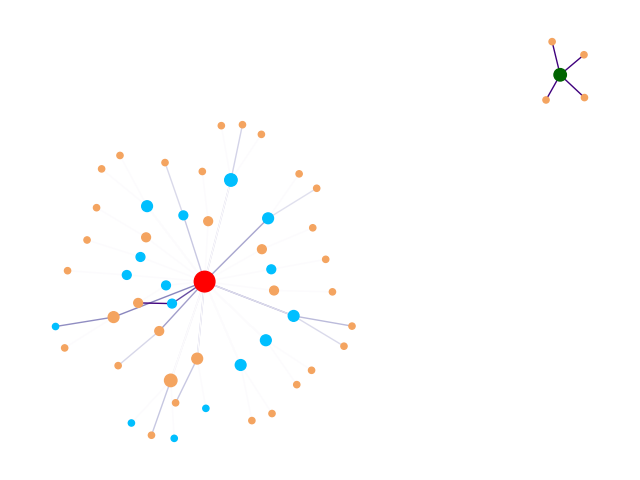}
        \label{fig:neg-co-case}
    }
    \subfigure[]{
        \includegraphics[width=0.4\linewidth]{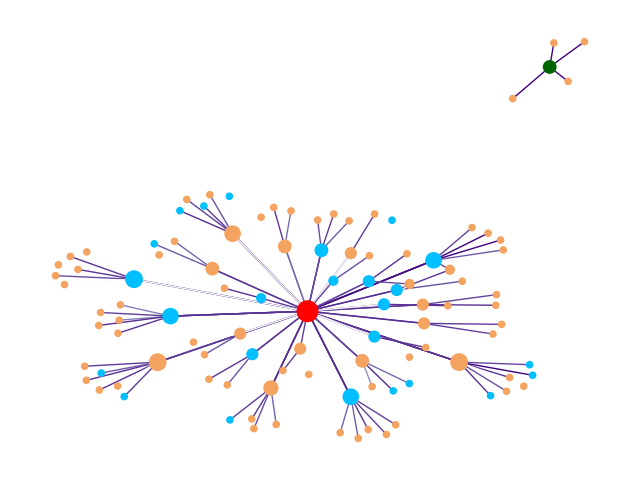}
        \label{fig:neg-soc-case}
    }
    \caption{Explanation graph visualization for case study. (a) and (b) are explanation graphs for positive sample in interaction tower and social tower respectively, while (c) and (d) are explanations for the low-ranking negative sample in the same tower order. For all four graphs, the orange and blue nodes represent user and item nodes, while the red, yellow and blackish green nodes represent target user, positive sample and negative sample respectively. Note that the blackish green negative sample nodes are in the upper-right corner of (c) and (d). The shade of edges in the user's ego-net represent the cosine similarity between their targeting nodes and corresponding candidate item.}
    \label{fig:case-study-graph}
\end{figure}

\begin{figure}
    \centering
    \subfigure[]{
        \includegraphics[width=0.16\linewidth]{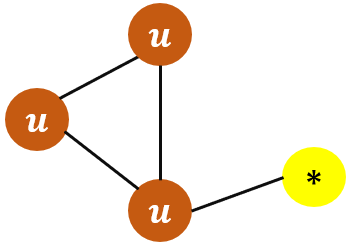}
    }
    \subfigure[]{
        \includegraphics[width=0.35\linewidth]{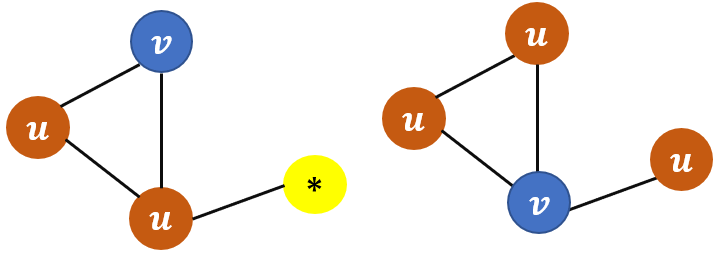}
    }
    \\
    \subfigure[]{
        \includegraphics[width=0.16\linewidth]{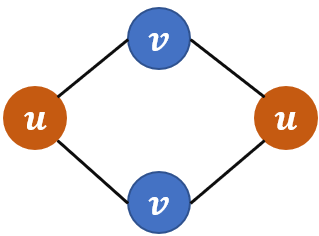}
    }
    \subfigure[]{
        \includegraphics[width=0.16\linewidth]{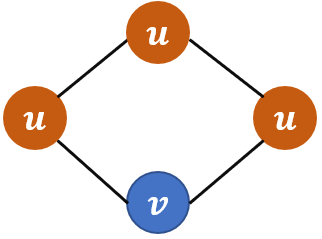}
    }
    \subfigure[]{
        \includegraphics[width=0.16\linewidth]{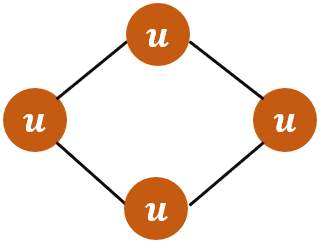}
    }
    \caption{Possible templates of triangles and quadrilaterals. (a)-(b) are two triangle types representing "friend of friend" ("fof" for short) relations in social network and co-purchase ("cop" for short) relations in interaction graph respectively. (c)-(e) are three quadrilateral types representing different combinations of "cop" and "fof" relations, indicating stronger connections among endpoint users.}
    \label{fig:motif-type}
\end{figure}

\begin{figure*}
    \centering
    \subfigure[Triangles in Yelp]{
        \includegraphics[width=0.22\linewidth]{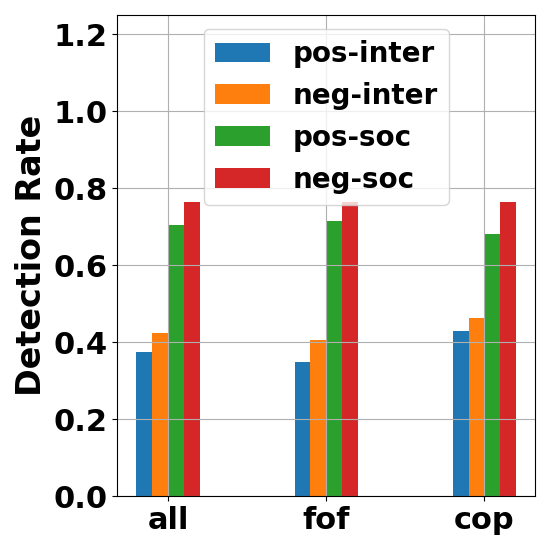}
    }
    \subfigure[Quads in Yelp]{
        \includegraphics[width=0.22\linewidth]{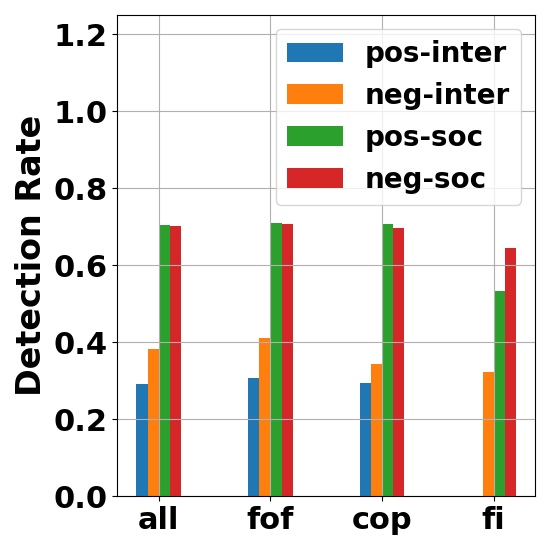}
    }
    \subfigure[Triangles in Ciao]{
        \includegraphics[width=0.22\linewidth]{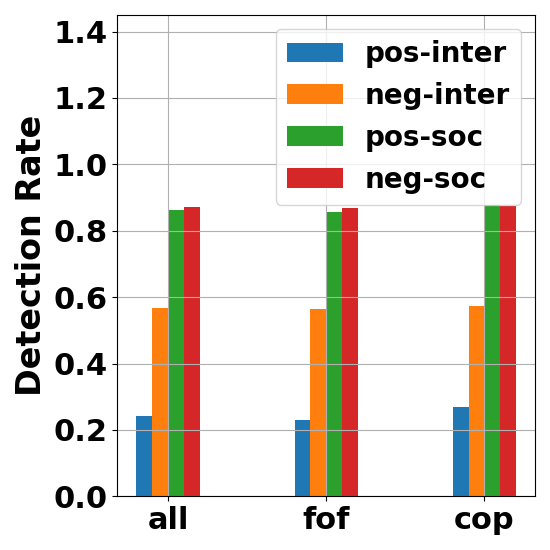}
    }
    \subfigure[Quads in Ciao]{
        \includegraphics[width=0.22\linewidth]{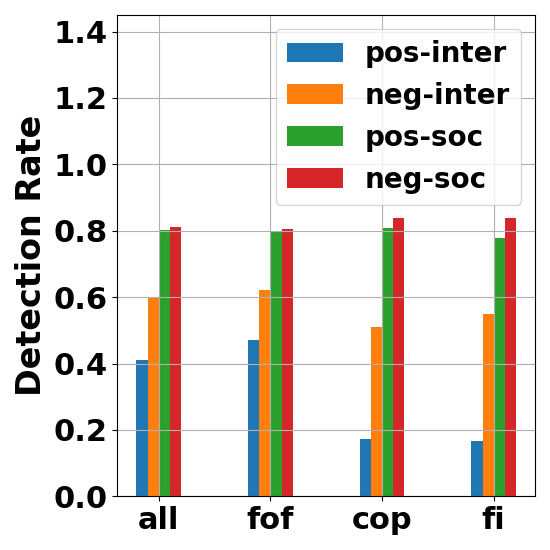}
    }
    \caption{Detection rate of triangles and quadrilaterals (quads for short) on Yelp and Ciao. The results are grouped by motif types, where "all" refers to motif-level statistics. Detailed data of different towers and positive/negative samples is also distinguished. Note that the fi motif in (b) is omitted, as there are no positive interactions for this type in the dataset.}
    \label{fig:detect-rate}
\end{figure*}

\begin{figure*}
    \centering
    \subfigure[Triangles in Yelp]{
        \includegraphics[width=0.22\linewidth]{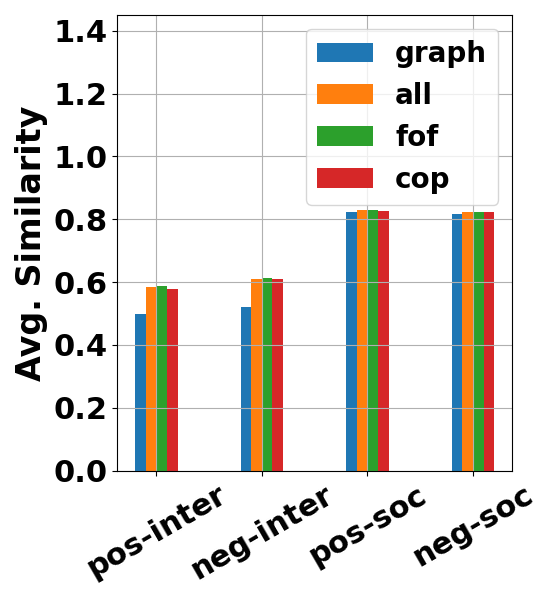}
    }
    \subfigure[Quads in Yelp]{
        \includegraphics[width=0.22\linewidth]{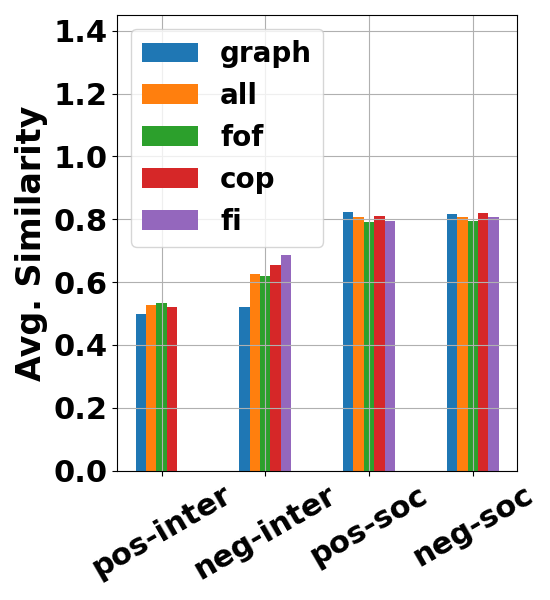}
    }
    \subfigure[Triangles in Ciao]{
        \includegraphics[width=0.22\linewidth]{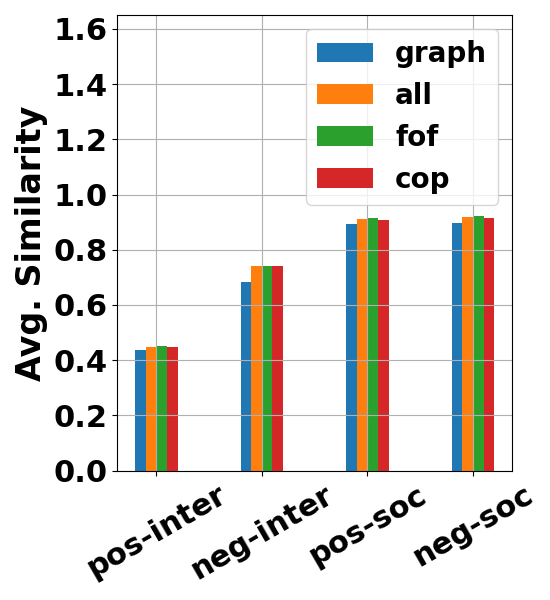}
    }
    \subfigure[Quads in Ciao]{
        \includegraphics[width=0.22\linewidth]{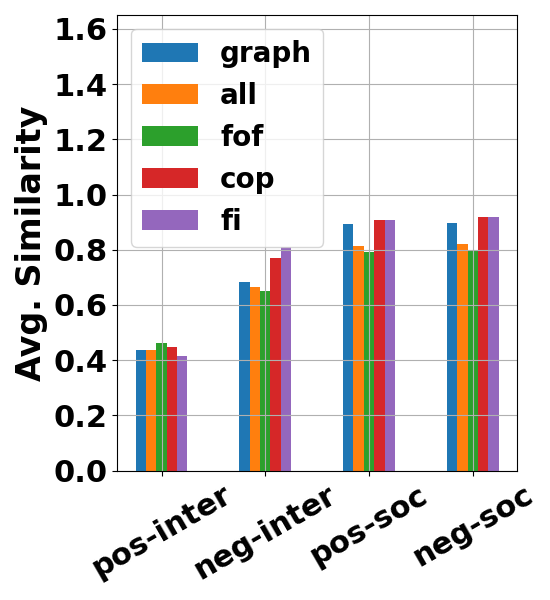}
    }
    \caption{Average similarity of triangles and quadrilaterals (quads for short) on Yelp and Ciao. The results are grouped by their tower and positive/negative sample belongings. Detailed data of different types of motifs is also presented. }
    \label{fig:importance-level}
\end{figure*}

\textbf{\textit{Case Study.}}
We conduct a case study to analyze the relations between explanatory subgraphs and ranking decisions. We select a representative test sample from Yelp dataset, visualizing the ego-paths activated by the positive sample and one of the low-ranking negative samples in each tower in Figure \ref{fig:case-study-graph}. We also visualize the 1-hop ego-nets of involved items to show the relations between candidates and candidate-aware explanations. We remove the repetitive ego-paths and reconnect ego-paths with the target user to be more intuitive. 
Following observations can be concluded: 
(1) Different towers have different focuses. Based on explanations for positive sample in Figure \ref{fig:pos-co-case} and \ref{fig:pos-soc-case}, we can find that the sampled ego-paths and their similarity distributions in different towers are quite different. In interaction tower, only a few ego-paths are of high similarity, and social relation based common neighborhood is paid more attention to. In contrast, more ego-paths are of high similarity in social tower, and more user-item interaction based common neighborhood is identified, which indicates that the user-item pair is closer from social perspective. 
(2) Items that have no common neighborhood with target user can also activate certain ego-paths as explanations. Based on negative sample related explanations in Figure \ref{fig:neg-co-case} and \ref{fig:neg-soc-case}, although there is no overlap between the neighborhood of negative sample and target user, we can still identify important ego-paths. 
(3) Comparative explanation can be made by directly comparing explanations of different samples. Based on the presented example, several paths can be found between target user and positive sample. The ego-paths are also of higher similarity with positive sample in both towers. In contrast, the selected ego-paths are less relevant with negative sample, and there is no common neighborhood between negative sample and target user. To this end, we have provided reasonable explanations for ranking the positive sample higher than the negative sample via comparison among explanation graphs.

Note that although the dense explanatory graphs make them not easily understandable for end users, our design aims to investigate comparative relationships. Dense explanatory graphs can provide detailed candidate-wise comparison, which can be more persuasive for both users and developers after further processing. 

\textbf{\textit{Systematic Pattern Analysis.}}
With ego-paths intertwined, various motifs can be formed. To demonstrate SoREX's ability for model-level explanation, we provide a toy example and analyze the systematic patterns of two simple motifs formed when two ego-paths intersect, namely triangles and quadrilaterals. Based on the number of user and item nodes in motifs, we can categorize them into two types of triangles and three types of quadrilaterals.
Figure \ref{fig:motif-type} presents templates and the real-world meanings of these detailed motif types. Considering that quadrilaterals are essentially two ego-paths with the same endpoints, we analyze quadrilaterals based on the types of ego-paths alternatively for simplicity. There are totally three kinds of templates for ego-paths, including "user-user-user" (friend of friend, "fof" for short), "user-item-user" (co-purchase, "cop" for short) and "user-user-item" (friend's interaction, "fi" for short). We leave more complex substructures for future work.
Our discussions will be extended from the perspective of detection rate and average similarity derived from Eq. (\ref{eq:path-sim}), which represent the quantity and importance of motifs respectively. 
Comparisons will also be made between two towers and between positive and negative samples to identify factor-specific and comparative patterns. 
Note that detection rate and average similarity relate to each other, because the similarity values directly determine the sampling probability.

We first analyze the detection rate of both triangles and quadrilaterals. 
Given motif type, detection rate measures the ratio of detected motifs to all motifs formed in $\hat{\mathcal{W}}_{ego}$. The motif-level and type-specific data on Yelp and Ciao are shown in Figure \ref{fig:detect-rate}. We have following observations: 
(1) In both datasets, detection rate of both triangles and quadrilaterals for positive sample related explanation graphs are less than the rate for negative samples related explanations. Considering that positive samples usually have common neighborhood with target users, the ego-paths connecting them will grab more attention when processing positive samples, while motifs with relational information will be more important for negative samples without common neighborhood.
(2) Detection rate in social tower is generally higher than which in interaction tower. 
(3) The detection rate of triangles is higher than quadrilaterals in Yelp, while the conclusion is opposite in Ciao. 
(4) In both datasets, interaction based "cop" triangles and social based "fof" quadrilaterals have higher detection rate in interaction tower for explanations of both positive and negative samples. There are no significant patterns in social tower.

Then we analyze the average similarities. We use the average similarity of all ego-paths involved in a motif as its similarity. The tower-specific data of Yelp and Ciao are shown in Figure \ref{fig:importance-level}. Following observations can be concluded: 
(1) In both datasets, the average similarities of all types of triangles are consistently higher than which of all ego-paths in both towers, indicating that triangles are important motifs in explanations. In contrast, the average similarities of quadrilaterals generally have no significant advantage over and may even be lower than the average level of explanation graphs. 
(2) Triangles are generally more important than quadrilaterals in both datasets.
(3) The average similarity distribution of different triangle types are close. In contrast, interaction-related "cop" and "fi" based quadrilaterals are generally of higher similarities compared with social-related "fof" based quadrilaterals in both tower.

\begin{table}[htbp]
  \centering
  \caption{NDCG@10-based fidelity scores across four datasets.}
    \begin{tabular}{c|cccc}
    \hline
    Method & Yelp  & Flickr & Ciao & LastFM \\
    \hline
    GSAT  & 2.37  & 7.88 & 9.53 & 3.18 \\
    SoREX w/ top-$K$ ego-path  & 2.75 & 8.13& 12.42& 4.01 \\
    SoREX w/ IB constraint  & 3.87& 8.91& 14.26&6.03\\
    SoREX w/o re-aggr & 3.95& 8.74& 15.29& 5.81\\
    SoREX  & \textbf{5.77}  & \textbf{10.38} & \textbf{20.83} & \textbf{7.27} \\
    \hline
    \end{tabular}%
  \label{tab:fidelity}%
\end{table}

\textbf{\textit{Quantitative Analysis.}}
We adopt fidelity scores to quantitatively demonstrate the effectiveness of our explanations. 
Fidelity score measures the performance drop when important features are removed. In our situation, if the model can capture more important ego-paths for each user-item pair, the removal of such ego-paths would lead to more significant performance drop. Considering that GSAT~\cite{gsat} only uses sampled explanatory subgraph for prediction just like our SoREX, we compare SoREX with GSAT and define the fidelity score based on NDCG score. Given user-item pair $(u_i, v_j)$, we first re-aggregate the ego-path subset $\hat{\mathcal{W}}_{ego}^* \backslash \tilde{\mathcal{W}}_{ego}^*(j)$ that are not sampled in each tower to obtain new predicted result $\hat{NDCG}_{ij}$. Then, we calculate the percentage of its performance drop compared with original NDCG@10 result $NDCG_{ij}$. Therefore, the fidelity score $\Delta \mathit{NDCG}\%$ is formulated as $\Delta \mathit{NDCG}\% = \frac{1}{|\mathcal{E}_{t}|}\sum_{(u_i,v_j)\in\mathcal{E}_t} \frac{(\mathit{NDCG}_{ij}-\hat{\mathit{NDCG}_{ij}})}{NDCG_{ij}}\%$. 
Similar procedures are also conducted for GSAT.
The results on all four datasets are listed in Table \ref{tab:fidelity}. Compared with GSAT, SoREX can learn factor-specific and candidate-aware ego-path similarity distributions, while GSAT assigns the same weight for each node regardless of candidate items. Meanwhile, without the re-aggregation module (w/o re-aggr), the model's fidelity scores decrease significantly, highlighting the critical role of the re-aggregation module in aligning interpretability with prediction results. In addition, the top-$K$ and IB constraint variants show further drops in fidelity, due to their hard truncation and information compression, respectively, both of which lead to loss of important explanatory signals. Thus, our SoREX can provide more appropriate and personalized explanations and consistently achieve higher fidelity scores.

\begin{figure}
    \centering
    \hspace{20pt}
    \subfigure[]{
    \hspace{-20pt}
        \includegraphics[width=0.45\linewidth]{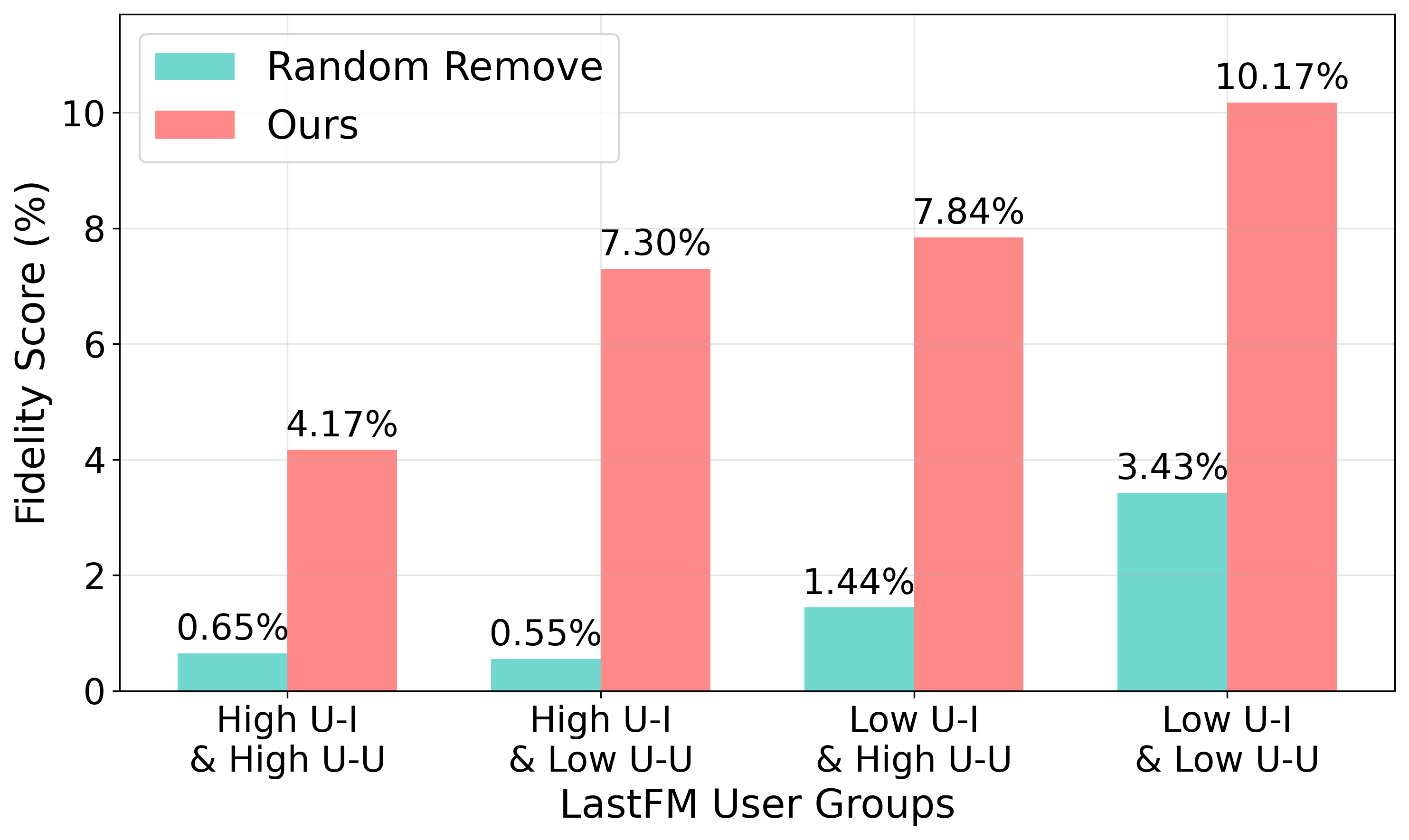}
    }
    \hspace{30pt}  
    \subfigure[]{
    \hspace{-20pt}
        \includegraphics[width=0.45\linewidth]{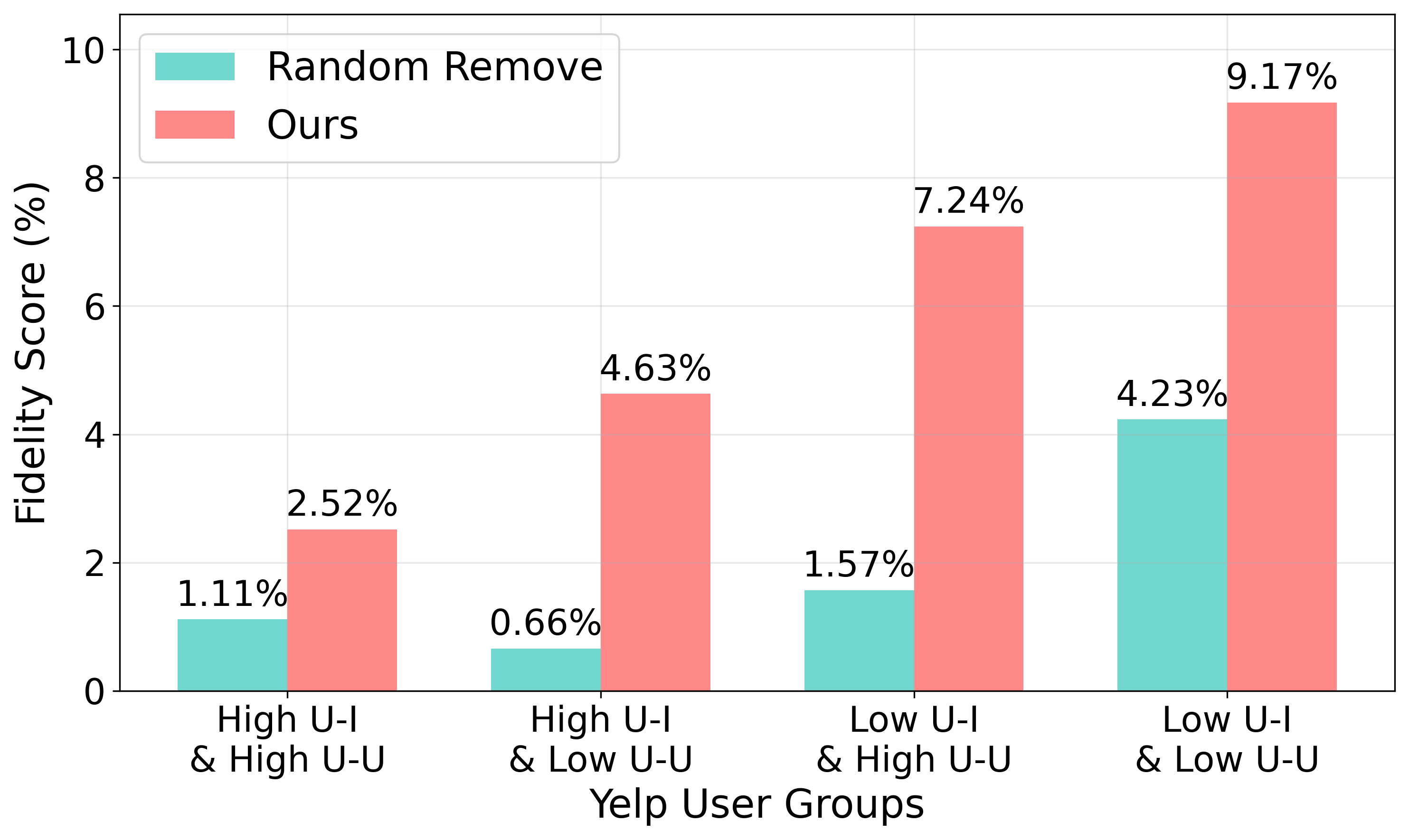}
    }
    \caption{Fidelity analysis across users with different characteristics. The fidelity of explanations is analyzed for user groups with different user–item and user–user degrees on Yelp and LastFM.}
    \label{fig:analyze}
\end{figure}

We further conduct a more in-depth analysis on the LastFM and Yelp datasets by examining the fidelity scores of users with different item graph and social graph degrees, in comparison with random path removal. Specifically, users are grouped into four categories based on whether their user–item (u–i) and user–user (u–u) degrees are above or below the median (i.e., 50th percentile), resulting in a $2\times2$ grouping: high u–i degree, low u–i degree, high u–u degree, and low u–u degree. As shown in Figure~\ref{fig:analyze}, for user groups with lower degrees, random removal of paths leads to a more significant drop in model performance. At the same time, our explanation paths consistently improve fidelity scores across all degree groups, with the benefit being most pronounced for users with low degrees. This suggests that for low-degree users, the identified ego-paths are particularly critical, as these users have fewer alternative paths. Consequently, providing faithful explanations for their recommendations is both more useful and more impactful.

\section{Limitations and Future Directions}

This work is an initial attempt to combine the concept of comparative explanation with graph-based social recommender explainer. Our proposed SoREX involves using relevant ego-path sampling to generate candidate-aware explanatory subgraphs. However, we believe there are still several challenges remaining:

\textit{Understandability.}
We propose to provide dense subgraph based explanations for detailed comparison among different candidate items. However, the graph-based explanations are not easy to understand, especially for those based on dense graphs. Although we could make the explanations more understandable by adopting certain post-processing programs or assigning real-world meanings to certain subgraph patterns, they are still not intuitive and precise enough. Balancing the precision and comprehensiveness of graph-based explanations remains an significant challenge. 

\textit{Scalability.} As we have analyzed in Section \ref{sec:complexity}, we compromise the time and memory complexity of SoREX to achieve better explainability. However, when the social recommenders are deployed in industrial million-scale graphs, such trade-off would be unacceptable. How to develop comparatively explainable social recommenders with sub-linear complexity remains a challenge. 

\textit{Evaluation of Explainer.} Although multiple evaluation methods are adopted for the evaluation of SoREX, the lack of ground truth explanatory subgraphs makes it difficult to perform further quantitative verification. Such lack of ground truth labels is not only the lack of labels themselves, but also the lack of proper label definitions. We could only explore some enlightening patterns based on certain subgraph templates, while the social recommendation explainers are highly likely to learn spurious correlations without the hints from external data. In fact, most social science related applications face similar limitations. Hence, we need an efficient and reliable strategy to evaluate self-explainable models without the access to ground truth explanations.

\section{Conclusion}

In this work, we introduce a novel self-explainable social recommendation framework SoREX to fill the explainability gap in GNN-based social recommendation. We first devise a social influence aware and friend recommendation enhanced two-tower framework to independently model the influence of social and interaction factors and lay the foundation for factor-specific explanation. Then, we propose to provide relevant ego-path based explanations. We transform ego-net of target user into a set of multi-hop ego-paths, and extract factor-specific and candidate-aware dense ego-path subsets for each candidate item, enabling comparative explanation among different candidates and factors via complex substructure investigation. We further perform explanation re-aggregation to emphasize relevant subgraphs of the user's ego-net and relate explanations to downstream predictions to make SoREX self-explainable. In addition, we design an auxiliary friend-recommendation task to capture more reliable friend relations and strengthen the social tower. Comprehensive experiments on four benchmark datasets demonstrate the effectiveness of SoREX in both predictive accuracy and explainability. 

\section*{Acknowledgments}
This research was supported by the Public Computing Cloud of Renmin University of China and by the Fund for Building World-Class Universities (Disciplines) at Renmin University of China.

\bibliographystyle{ACM-Reference-Format} 
\bibliography{sample-base}

\end{document}